\IfFileExists{leer.eps}{}{
}
\let\oldvec\vec
\documentclass[epj]{svjour}
\let\vec\oldvec 

\usepackage[intlimits]{amsmath}
\usepackage{graphicx}
\usepackage[colorlinks=true,linkcolor=blue,citecolor=blue,urlcolor=blue]{hyperref}
\usepackage{multirow}
\usepackage{slashed}
\usepackage[usenames,table]{xcolor}
\usepackage[numbers,sort&compress]{natbib}

\newcommand{\ONE}{ \mathbf{1} } 

\newcommand{\Tr}{\mathrm{Tr}}
\newcommand{\ii}{\mathrm{i}}
\newcommand{\Rpos}{\mathtt{R}}
\begin{document}\sloppy

\title{The quark-gluon vertex in Landau gauge bound-state studies}

\author{Richard Williams\thanks{\email{richard.williams@theo.physik.uni-giessen.de}}}

\institute{Institute of Theoretical Physics, Justus-Liebig University of Giessen, \\ Heinrich-Buff-Ring 16, 35392, Giessen, Germany}  

\date{Received: date / Revised version: date}

\abstract{
We present a practical method for the solution of the quark-gluon vertex 
for use in Bethe--Salpeter and Dyson--Schwinger calculations. The 
efficient decomposition into the necessary covariants is detailed, with 
the numerical algorithm outlined for both real and complex Euclidean momenta.
A truncation of the quark-gluon vertex, that neglects explicit back-coupling to enable
the application to bound-state calculations, is given together with
results for the quark propagator and quark-gluon vertex for different
quark flavours. The relative impact of the various components of the 
quark-gluon vertex is highlighted with the flavour dependence of the 
effective quark-gluon interaction obtained, thus providing insight 
for the construction of phenomenological models within Rainbow-Ladder. 
Finally, we solve the corresponding Green's functions for complex
Euclidean momenta as required in future bound-state calculations.
\PACS{
      {11.10.St}{}\and
      {11.15.-q}{}\and
      {11.30.Rd}{}\and
      {12.38.Lg}{}
}
}

\maketitle

\section{Introduction}\label{sec:introduction}
The phenomena of QCD such as confinement and dynamical chiral symmetry breaking (DCSB) are exemplified through the spectrum of hadrons, their decays, and transition form factors. Thus, a non-perturbative description of bound-states in terms of the component quarks and gluons is required. 
One such approach are the Dyson--Schwinger (DSE) and Bethe--Salpeter (BSE) equations which provide a framework where chiral properties -- in particular the Goldstone nature of pions -- can be taken into account through constraints such as the axial-vector Ward--Takahashi identity (axWTI)~\cite{Munczek:1994zz,Bender:1996bb} amongst others~\cite{He:2000we,Pennington:2005mw,Qin:2013mta}. 

Since the DSEs constitute an infinite tower of coupled non-linear integral equations, truncations must be employed to make their solution tractable; for recent reviews, see Refs.~\cite{Roberts:1994dr,Alkofer:2000wg,Maris:2003vk,Fischer:2006ub,Bashir:2012fs,Roberts:2012sv}.  These can be performed at varying levels of sophistication, with the simplest being that of Rainbow-Ladder (RL) wherein only the DSE for the quark is directly considered and the coupling between quarks and gluons is given by an effective interaction. This reduces the quark-(anti)quark interaction to a simple flavour independent coupling that is a capable model of DCSB.  Such a truncation has yielded much phenomenological success for mesons~\cite{Maris:1997tm,Maris:1999nt,Maris:1999bh,Jarecke:2002xd,Maris:2002mz,Maris:2003vk,Maris:2005tt,Krassnigg:2009zh,Krassnigg:2010mh,Chang:2013nia} as well as baryons~\cite{Eichmann:2009qa,Eichmann:2011vu,Eichmann:2011pv,SanchisAlepuz:2011jn,Roberts:2011cf,Segovia:2013uga}.

However, the main detriment of RL is that the structure is limited to that of a $\gamma^\mu\otimes\gamma^\mu$ interaction with no variation in the projected strength; to remedy this one must go beyond RL~\cite{Bender:1996bb,Watson:2004kd,Bhagwat:2004hn,Fischer:2005en,Maris:2005tt,Matevosyan:2006bk,Fischer:2008wy,Alkofer:2008et,Fischer:2009jm,Chang:2009zb,Bashir:2012fs,Heupel:2014ina,Vujinovic:2014ioa}.  
This is fairly straightforward in principle since one knows the gluon propagator very well from both Lattice studies and functional methods~\cite{Cornwall:1981zr,Watson:2001yv,Lerche:2002ep,Zwanziger:2002ia,Pawlowski:2003hq,Alkofer:2004it,Fischer:2006vf,Aguilar:2008xm,Fischer:2009tn}. This leaves the quark-gluon vertex as the central object in non-perturbative and
bound state studies of hadrons, for it explicitly connects the matter sector to the gauge sector. It is known that the enhancement that triggers DCSB is provided by this vertex~\cite{Alkofer:2006gz,Alkofer:2008tt}. Moreover,
other non-perturbative effects such as pion-cloud corrections are contained within~\cite{Watson:2004jq,Thomas:2008bd,Fischer:2007ze,Fischer:2008wy,Heupel:2013zka}, together with the dominant dependence on quark flavour necessary to make a connection to the heavy quark 
limit~\cite{Maris:2006ea,Popovici:2010mb,Popovici:2011yz,Blank:2011ha}. Whilst early studies of the quark-gluon vertex within Lattice QCD exist~\cite{Skullerud:2002ge,Skullerud:2003qu,Kizilersu:2006et}, an updated calculation is required before constraints or quantitative comparisons can be made.

\begin{figure}[t]
\centering{\resizebox{0.20\textwidth}{!}{\includegraphics{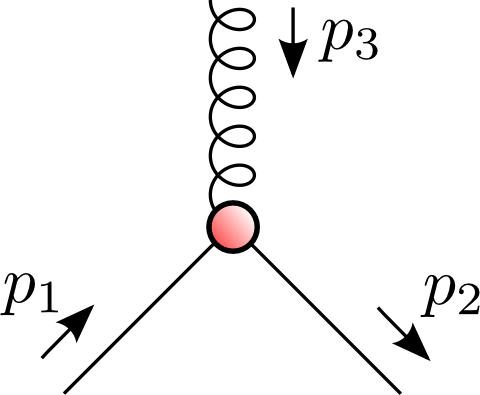}}}
\caption{The dressed quark-gluon vertex.}
\label{fig:qgv-momenta}
\end{figure}

In this paper, we present a solution strategy for a class of truncations of the quark-gluon vertex and quark propagator for both space-like~\cite{Alkofer:2008tt,Aguilar:2014lha,Hopfer:2013np,Windisch:2012de,Rojas:2013tza} Euclidean momenta and their analytic continuation to time-like momenta needed in bound-state 
equations~\cite{Fischer:2009jm,Williams:2009wx}.  
We provide a specific model, constrained by the requirement that it be applicable for future bound-state studies, and investigate the flavour dependence of the quark propagators and quark-gluon vertex.

\section{Quark-gluon vertex}

\begin{figure*}[!t]
\centering{\resizebox{0.7\textwidth}{!}{\includegraphics{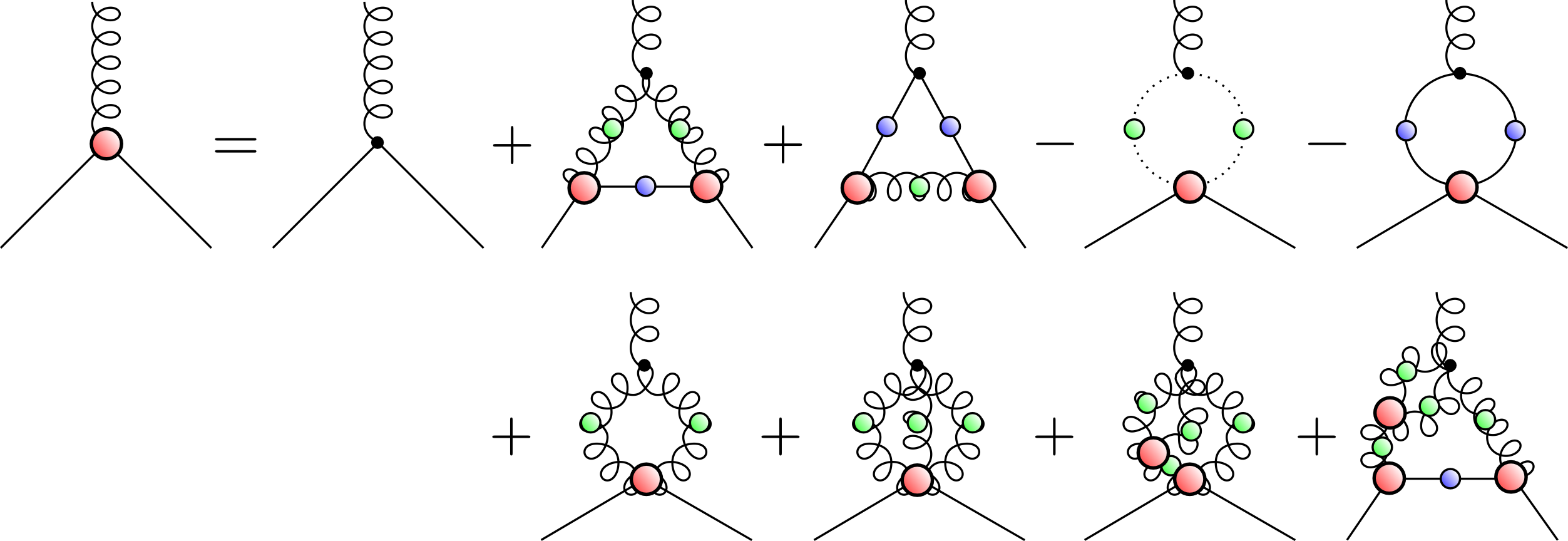}}}
\vspace{0.5cm}\\
\centering{\resizebox{0.6\textwidth}{!}{\includegraphics{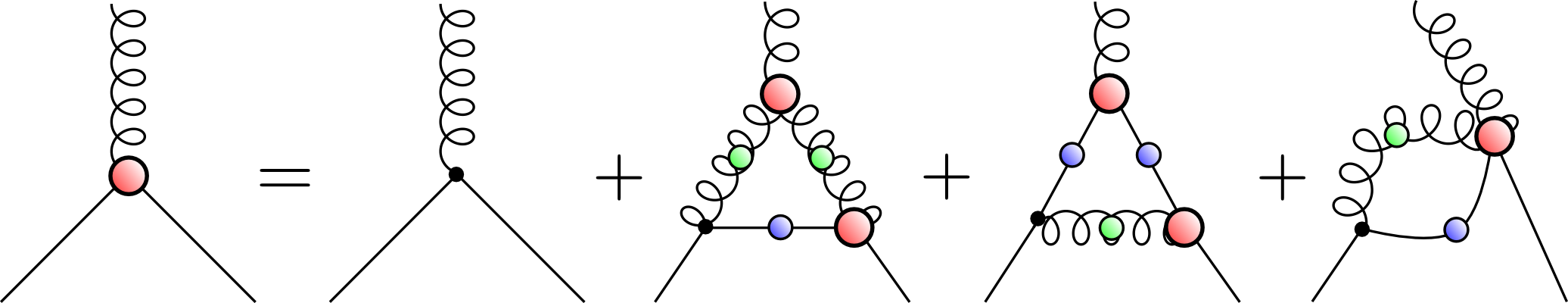}}}
\caption{First (top) and second (bottom) forms of the DSE for the quark-gluon vertex. All internal propagators are dressed, with filled circles indicating dressed vertices. Other quantities are bare.\label{fig:qgvdse}}
\end{figure*}

\begin{figure}[!b]
\centering{
\resizebox{0.95\columnwidth}{!}{\includegraphics{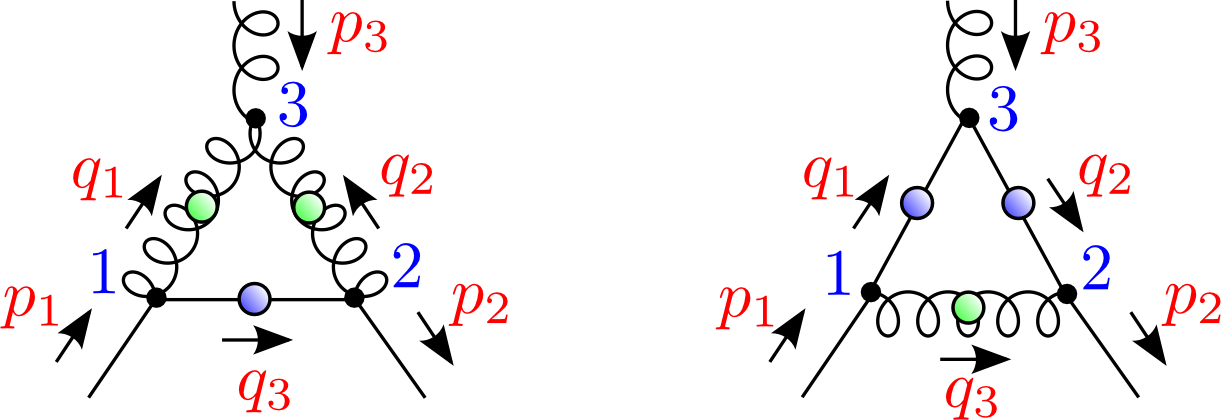}}}
\caption{Non-Abelian (left) and Abelian (right) classes of a diagram considered in our truncation of the quark-gluon vertex. Any combination of the internal vertices $1$, $2$, $3$, labelled counter-clockwise, may be dressed depending upon the truncation at hand.}\label{fig:naab}
\end{figure}

We employ herein the colour reduced quark-gluon vertex $\Gamma^{a}_\mu (l,k) = t^a\, \Gamma_\mu(l,k)$
where $k=p_3$ is the incoming gluon momentum and $l=\left(p_1+p_2\right)/2$ the relative quark momentum, as given in Fig.~\ref{fig:qgv-momenta}. 
This object is obtained through solution of its corresponding DSE, of which two forms are given in Fig.~\ref{fig:qgvdse} with the difference being to which external leg the bare vertex is attached.
Both equations are functions also of higher $n$-point functions and must be truncated to enable solution. 
In performing such a truncation we must find a balance between its numerical or algebraic complexity and the physics that it contains. In the case of the quark-gluon vertex, it is sensible to neglect the two-loop diagrams, and to either eliminate non-primitively divergent $n$-point functions or replace them via a dressed skeleton expansion~\cite{Alkofer:2008jy,Alkofer:2008tt}. The remaining diagrams can then be ordered by a non-perturbative counting scheme such as large $N_c$. Alternatively, for \emph{scaling} solutions one can order the diagrams according to infrared power counting~\cite{Fischer:2006vf,Fischer:2009tn}, and assume that the hierarchy is transferred parametrically to the decoupling scenario considered here.   The impact of neglecting various contributions can be mitigated in part by effectively dressing previously bare vertices through a re-ordering of the diagrammatic resummation. 
Typically, in a first calculation, one considers the two contributions given in Fig.~\ref{fig:naab} with any combination of bare or dressed internal vertices, labelled $1$ through $3$~\cite{Alkofer:2006gz,Alkofer:2008tt}. For example, with all internal vertices dressed one would have a truncation reminiscent of the $3$PI formalism~\cite{Berges:2004pu}, $2$ and $3$ dressed would match the second form of the DSE for the quark-gluon vertex, and finally with $1$ and $2$ dressed the first form.

We make use of transverse and/or orthonormal momenta~\cite{Eichmann:2011vu,Eichmann:2012mp} which have proven efficient in the covariant investigation of baryons~\cite{Eichmann:2009qa,Nicmorus:2010sd,SanchisAlepuz:2011aa,SanchisAlepuz:2011jn,Sanchis-Alepuz:2013iia}. In particular, the strategy is optimised such that the injection of a complex momentum (required for the analytic continuation of Euclidean momenta to the time-like region as probed in bound-state studies) can be easily considered.  Numerical solutions are performed using the shell-method, as described in Section~\ref{sec:ancont}.

In general the vertex may be decomposed into twelve components 
\begin{align}
\Gamma^\mu(l,k) = \sum_{i=1}^4\sum_{a=1}^3c_{a}^{i}(l,k) L^\mu_{(a)}R_{(a)}^i\;,
\end{align}
where $c_{a}^{i}(l,k)$ are scalar coefficients that parametrise the Lorentz ($L^\mu_{(a)}$) and 
Dirac ($R_{(a)}^i$) parts of the basis (see Appendix ~\ref{sec:dsetrans}--\ref{sec:dsenontrans} for details). For convenience, we
collapse the indices $i, a$ and write
\begin{align}\label{eqn:collapsed indices1}
T_{1\ldots4}  = c^{1\ldots4}_{(1)}\;,\;\;\;
T_{5\ldots8}  = c^{1\ldots4}_{(2)}\;,\;\;\;
T_{9\ldots12} = c^{1\ldots4}_{(3)}\;,
\end{align}
where note that the indices $a$ may be referred to by symbols in the sequel.

Generically, the form of the DSE for the quark-gluon vertex is given by summation of self-energy contributions
\begin{align}
\Gamma^\mu(l,k) = Z_{1f}\gamma^\mu + \Lambda^\mu_{NA} + \Lambda^\mu_{AB} +\ldots
\end{align}
where $\Lambda^\mu_{NA}$, $\Lambda^\mu_{AB}$ are representative corrections that
we consider later and the ellipsis refers to other terms present in the DSE, see Fig.~\ref{fig:qgvdse}. Projection
onto the coefficients is obtained via
\begin{align}\label{eqn:coefficientwithtreelevel}
c_{(a)}^{i}(l,k) &= \bar{L}^\mu_{(a)} \Tr\left[ \bar{R}^i_{(a)} \gamma^\mu\right] +\bar{L}^\mu_{(a)} \Tr\left[ \bar{R}^i_{(a)}\Lambda^\mu_{\textrm{NA}}\right]\nonumber \\
&+\bar{L}^\mu_{(a)} \Tr\left[ \bar{R}^i_{(a)}\Lambda^\mu_{\textrm{AB}}\right] + \ldots\;.
\end{align}
In an obvious notation, we write the contribution to the coefficient $c_{(a)}^i$ of each self-energy
contribution as $c_{(a)}^{i,\,\textrm{NA}}$ and $c_{(a)}^{i,\,\textrm{AB}}$ respectively.

\subsection{Non-Abelian diagram}
Let us apply the transverse basis decomposition of Section~\ref{sec:dsetrans} to the case of the (\emph{a posteriori}) dominant non-Abelian contribution
to the quark-gluon vertex. From here on we focus only on the Dirac and
Lorentz structure of the diagram, given in the left panel of Fig.~\ref{fig:naab}, and implicitly contract all colour indices and suppress the renormalisation factors. The eventual choice of whether an internal vertex is taken dressed or not is left to the sophistication of the desired truncation, but note for consistency that each bare vertex must be associated with the corresponding multiplicative renormalisation factor.

Consider the following choice of four-momenta
\begin{eqnarray}
k^\mu =& p_3^\mu&= \left|k\right|\,(\begin{array}{cccc} 0, & 0, & 0, & 1 \end{array})\,, \\
l^\mu =& \left(p_1^\mu+p_2^\mu\right)/2 &= \left|l\right|\,(\begin{array}{cccc} 0, & 0, & z^\prime, & z \end{array}) \,,
\end{eqnarray}
where $z=\hat{k}\cdot \hat{l}$ and $z^\prime = \sqrt{1-z^2}$.
The non-Abelian diagram of Fig.~\ref{fig:naab} would have the following form
\begin{align}
\Lambda^\mu_{\textrm{NA}}(l,k)&=\frac{N_c}{2}\!\!\int\!\!\! \frac{d^4q}{\left(2\pi\right)^4}\Gamma_\alpha\left(l_1,-q_1\right) S(q_3) \Gamma_\beta\left(l_2,-q_2\right)\nonumber \\
&\qquad\times\, \Gamma^{\alpha^\prime\beta^\prime\mu}_{3g}(q_1,q_2,p_3) D^{\alpha\alpha^\prime}(q_1)D^{\beta\beta^\prime}(q_2)\;,
\end{align}
where the relative quark momenta of the internal vertices are defined $l_1^\mu = (p_1+q_3)^\mu/2$, $l_2^\mu = (p_2+q_3)^\mu/2$, and $q$ is the Euclidean loop momentum of which $q_i$ are appropriate functions. The $N_c/2$ stems from the colour trace. This integrand can be separated into a Dirac and Lorentz part
\begin{align}
M^{\alpha\beta}_{(a)} &= \bar{L}^\mu_{(a)}\, \Gamma_{3g}^{\alpha^\prime\beta^\prime\mu}(q_1,q_2,p_3)\,D^{\alpha\alpha^\prime}(q_1)\,D^{\beta\beta^\prime}(q_2)\;, \label{eqn:MAB} \\
N_{(a)\,\alpha\beta}^{i}&=\Tr\left[\bar{R}_{(a)}^i \Gamma_\alpha(l_1,-q_1) S(q_3) \Gamma_\beta(l_2,-q_2)\right]\;,
\end{align}
where $a=(v,r,s)$, $i=1,\ldots,4$, and the contribution to the coefficients of the basis, Eq.~(\ref{eqn:basisprojection1}), is obtained through projection
\begin{align}
c_{(a)}^{i,\,\textrm{NA}}(l,k) =  \frac{N_C}{2}\int \frac{d^4q}{\left(2\pi\right)^4} M_{(a)}^{\alpha\beta} N^i_{(a)\,\alpha\beta}\,.
\end{align}
Note that $a$ is not summed over here and merely labels the basis components.

Though at first sight this appears to be an unnecessary separation, in fact it allows for the
systematic construction of truncation schemes to be tackled more easily. 
For example, let us assume that tree-level quark-gluon vertices are employed under
the loop integral, together with a tree-level three-gluon vertex. On replacing this with, say,
a dressed three-gluon vertex~\cite{Cucchieri:2008qm,Dudal:2012zx,Huber:2012kd,Binosi:2013rba,Aguilar:2013vaa,Blum:2014gna,Eichmann:2014xya}, only $M^{\mu\nu}_{(a)}$ would be modified. Similarly,
any changes to the internal quark-gluon vertices affect only $N_{(a)\,\mu\nu}^i$. In addition,
since the Lorentz part $M_{(a)}^{\mu\nu}$ is typically independent of the quark propagator 
and quark-gluon vertices it can be  pre-computed and stored.
This is especially useful when employing traditional computer algebra methods to trace
out the algebra, since the explicit contraction of the Dirac part with the Lorentz
part tends to lead to large algebraic expressions.

\subsection{Abelian diagram}
The contribution often considered to be sub-leading~\footnote{This is not true in all kinematic limits; see for example Refs.\cite{Davydychev:2000rt,Chang:2010hb} for a discussion of the the quark anomalous magnetic moment where this diagram dominates.}, due to it being colour suppressed by $N_c^2$, is the so-called Abelian diagram given in the right panel of Fig.~\ref{fig:naab}. Considering the same choice of external momenta as before, this diagram can be written
\begin{align}
\Lambda_{\textrm{AB}}(l,k)^\mu&=\frac{-1}{2N_c}\int\!\! \frac{d^4q}{\left(2\pi\right)^4}
\Gamma_\alpha\left(l_1,-q_3\right) 
S(q_1)
\Gamma_\mu\left(l_2,p_3\right) 
 \nonumber \\
&\qquad\qquad \times S(q_2)\Gamma_\beta\left(l_3, q_3\right)
D^{\alpha\beta}(q_3)\,,
\end{align}
where $l_1^\mu = (q_1+p_1)^\mu/2$, $l_2^\mu = (q_1+q_2)^\mu/2$, $l_3^\mu= (p_2+q_2)^\mu$ are the relative momenta of the internal
vertices. The Euclidean loop momentum is $q$, of which $q_i$ are appropriate functions. 
In the context of BSE studies, it has been investigated in 
Refs.~\cite{Bender:1996bb,Bender:2002as,Bhagwat:2004hn,Watson:2004kd,Matevosyan:2006bk,Alkofer:2008tt,Williams:2009wx}.

The Abelian contribution to the quark-gluon vertex does not benefit from the same
separation of the Dirac part from the Lorentz part. Instead we find that
\begin{align}
c_{(a)}^{i,\,\textrm{AB}}(l,k) &= \frac{-1}{2N_c}\int \frac{d^4q}{\left(2\pi\right)^4}  D^{\alpha\beta}(q_3)  \\  
&\!\!\!\!\!\!\!\!\Tr \bigg[\bar{R}_{(a)}^i \Gamma_\alpha\left(l_1,-q_3\right) \left(\bar{L}_{(a)}^\mu\chi_\mu\left(l_2,p_3\right) \right)
\Gamma_\beta\left(l_3, q_3\right) \bigg] \;,\nonumber
\end{align}
which is the Lorentz contraction and Dirac trace applied to the single spin-line. Note that we used
\begin{align}
S(q_1)\, \Gamma_\mu(l_2,p_3)\, S(q_2) := \chi_\mu(l_2,p_3)\;,
\end{align}
with $\Gamma$ now analogous to the Bethe-Salpeter amplitude and $\chi$ its wavefunction. This 
provides significant simplification of the algebra in the case that the top most vertex is dressed, and permits solution as an inhomogeneous BSE wherein the inhomogeneous term would be the non-Abelian diagram plus the tree-level term.

Note that up to a colour factor and the additional inclusion of crossed-ladder contributions, the Abelian diagram with the $2+3$ vertices dressed was solved inhomogeneously in Ref.~\cite{Williams:2009wx} in the
context of the bound-state for a vector meson. Such methods can be similarly applied here also in the complex plane.

\subsection{Analytic continuation}\label{sec:ancont}
As indicated earlier, bound-state calculations require that the quark propagator must be analytically continued to a bounded parabolic region of the complex plane. 

Starting with Eq.~(\ref{eqn:quarkdse}), one would ordinarily exploit $O(4)$ symmetry and take the incoming momentum to be $p^\mu = \left|p\right| (0,0,0,1)$. For its analytic continuation, we inject a complex momentum
$P^\mu = \ii\,M\,(0,0,0,1)$ parallel to $p$, so that the incoming (viz. outgoing) quark
momentum $p^{\prime\mu}$ is $(p+P/2)^\mu$. It is obvious that the corresponding squared invariant
$p^{\prime2}=\left(p+P/2\right)^2$ then maps out a bounded parabolic region of the complex plane with
vertex $-M^2/4$. Thus the heavier the bound-state the larger the region in the complex plane that needs to be explored.

We work in Euclidean space where the integration measure is space-like and we are free
to choose the momentum routing in the self-energy diagram of the quark DSE. For example, we can pass the 
momentum $(p^\prime-k)$ through the gluon and the loop momentum $k$ through the internal quark propagator. Then, solution of the quark
propagator in the complex plane requires the quark propagator on the
positive real axis together with a prescription for the gluon and quark-gluon
vertex (perhaps given by Ansatz in \emph{e.g.} the RL truncation) valid for complex momenta. Such an
approach requires no iteration of the quark DSE provided the quark propagator is
known for $p^{\prime2}\in\Rpos^+$~\cite{Maris:1997hd,Alkofer:2002bp}. However, the obvious caveat is that typically
we do \emph{not} know the gluon or the quark-gluon vertex
in the complex plane (exceptions are~\cite{Strauss:2012dg,Fischer:2009jm}). Moreover, solution of the quark-gluon vertex may require the three-gluon
vertex as input which has only been investigated for real momenta~\cite{Huber:2012kd,Huber:2012zj,Aguilar:2013vaa,Blum:2014gna,Eichmann:2014xya}.

The solution to this problem is to choose an alternative momentum routing so that
the complex momentum $p^\prime-k$ passes through the internal quark propagator, and the real Euclidean loop momentum $k$ through the gluon propagator. Since the parabolic regions are nested, we can expand out from the real-axis in parabolic
shells, iterating for each until convergence is achieved and then proceeding
outwards. Such a process is dubbed the shell-method~\cite{Fischer:2008sp} although other similar techniques exist~\cite{GimenoSegovia:2008sx,Krassnigg:2009gd,Windisch:2013dxa}. Then, the momentum passing
through the gluon is real and only the relative quark-momentum of the
quark-gluon vertex is continued to the complex plane. Through a judicious choice of the momenta
we can again pass this complex
momentum through the internal quark-line only. Thus, the quark-gluon vertex
can also be solved for expanding parabolic shells. This necessitates a micro/macro iteration that is repeated until mutual convergence.

\section{Ingredients}

\subsection{Quark propagator}
The Green's function fundamental to our discussion of mesons is that of the quark
propagator, since without it we cannot describe the matter fields that form
colorless bound-states. In its relatively simple structure are encoded such
non-perturbative properties as the dynamical generation of mass and the realisation
of a non-zero vacuum condensate through DCSB.
Moreover, chiral symmetry as expressed through the 
axWTI connects the interaction part of the quark
DSE to the quark-(anti)quark scattering
kernel required in the covariant description of bound states.

\begin{figure}[t]
\centering{\resizebox{0.95\columnwidth}{!}{\includegraphics{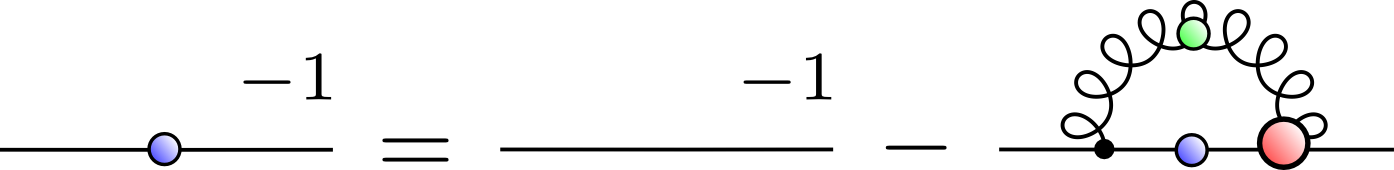}}}
\caption{The DSE for the fully dressed quark
propagator. Springs represent gluons and straight lines quarks. Large
filled circles indicate the quantity is fully-dressed, otherwise it is bare.}\label{fig:quarkdse}
\end{figure}

This DSE for the (inverse) quark propagator, shown in 
Fig.~\ref{fig:quarkdse}, is
\begin{align}\label{eqn:quarkdse}
  S(p)^{-1} &= Z_2 S_0^{-1}(p)\\
            &- g^2C_F\, Z_{1f}\!\int_k\gamma_\mu S(p-k)\Gamma_\nu(l,k)D_{\mu\nu}(k)\;, \nonumber
\end{align}
with $l=\left(p-k/2\right)$ the relative quark momentum of the internal vertex and $\int_k=\int d^4k/\left(2\pi\right)^4$. The fully-dressed inverse quark-propagator is diagonal in colour space
\begin{equation}
S^{-1}(p) = -i \slashed{p} A(p^2) + \ONE B(p^2)\;,
\end{equation}
and its bare counterpart given by $S^{-1}_0(p) = -ip\!\!\!\!\!\!\not~ +Z_m m $. Here,
$Z_2$, $Z_m$ and $Z_{1f}=Z_2/\tilde{Z}_3$ are respectively the renormalisation constants of the 
quark field, quark mass and quark-gluon vertex in Landau gauge using the miniMOM scheme~\cite{vonSmekal:2009ae}; $\tilde{Z}_3$ renormalises the ghost propagator. We deal with QCD where the number of colors is 3, and so
the Casimir $C_F=4/3$. The propagator is parametrised by two scalar functions $A(p^2)=1/Z_f(p^2)$ and $B(p^2)=M(p^2)/Z_f(p^2)$
where $Z_f$ is the quark wavefunction and $M$ is the quark mass function. $D_{\mu\nu}(k)$ is the gluon propagator in Landau gauge.

The solution of this system is therefore contingent upon two inputs. First is the aforementioned gluon propagator. Second, to close the system, we need the quark-gluon vertex.

\subsection{Gluon propagator}
The gluon propagator is given by
\begin{align}\label{eqn:gluonprop}
D_{\mu\nu}(k) = T^{(k)}_{\mu\nu}\, \frac{Z(k^2)}{k^2}\;,
\end{align}
where $Z(k^2)$ is the gluon dressing function and the transverse projector is $T^{(k)}_{\mu\nu} = \delta_{\mu\nu}-k_\mu k_\nu/k^2$. 
The employed ghost and gluon propagators come with associated renormalisation constants $Z_3$ and $\tilde{Z}_3$, together with a particular choice of the strong coupling constant $g_s^2$ such that the running of the ghost-gluon vertex is given in physical units. We use solutions in the decoupling scenario, see Ref.~\cite{Fischer:2008uz,Eichmann:2014xya} for details. For consistency, and to ensure that the reproduced anomalous dimensions are correct, all other renormalisation constants are appropriately related via their Slavnov-Taylor identities. This is necessary to obtain coincidence with the running coupling associated with each primitively divergent vertex at perturbative momenta.

\subsection{Three-gluon vertex}
The three-gluon vertex in pure Yang-Mills has recently been the focus of several investigations. Its tree-level structure can be
directly read off from the QCD Lagrangian, giving the reduced tree-level vertex
\begin{equation}\label{eqn:treelevel3gvertex}
\Gamma_{\textrm{3g}}^{(0)\mu_1\mu_2\mu_3}(p_1,p_2,p_3) =
  \delta^{\mu_1\mu_2} \left( p_1-p_2\right)^{\mu_3} 
+\textrm{cyclic} \;,
\end{equation}
where the coupling constant and colour factor $gf^{abc}$ are factored out.
There are
$10$ longitudinal components and $4$ transverse components, yielding $14$
components in total. Examples of such a basis are given in Refs.~\cite{Kim:1979ep,Ball:1980ax,Aguilar:2013vaa}.
In Landau gauge QCD it is sufficient retain only the transverse terms; a complete basis
is presented in Ref.~\cite{Eichmann:2014xya}. There, it was determined that the transversely projected tree-level
component is dominant, with other structures contributing at the $10\%$ level. Hence, here we keep only 
the tree-level component of the three-gluon vertex, as projected onto this transverse basis.
\begin{align}\label{eqn:3gvertexdress}
\Gamma^{\mu_1\mu_2\mu_3}_{3g,TTT}(p_1,p_2,p_3) &= F_1(p_1,p_2,p_3) \nonumber \\
&\times T^{(p_1)}_{\mu_1\mu_1^\prime}T^{(p_2)}_{\mu_2\mu_2^\prime}T^{(p_3)}_{\mu_3\mu_3^\prime}
\Gamma^{(0)\mu_1^\prime\mu_2^\prime\mu_3^\prime}_{3g}\;,
\end{align}
where $F_1(p_1,p_2,p_3)$ is a scalar function parametrising the leading component of the three-gluon dressing; to good approximation this can described by the single combination $\mathcal{S}_0=\left(p_1^2+p_2^2+p_3^2\right)/6$.
Since the basis is transverse, the transverse projector contained within the gluon propagator, Eq.~(\ref{eqn:gluonprop}), is redundant and the Lorentz part of the non-Abelian contribution to the quark-gluon vertex, Eq.~(\ref{eqn:MAB}), reduces to
\begin{align}
M^{\alpha\beta}_{(a)} &= \left(\bar{L}^\mu_{(a)}\, \Gamma_{3g,TTT}^{\alpha\beta\mu}\right)\, \frac{Z(q_1^2)}{q_1^2}\frac{Z(q_2^2)}{q_2^2}\;,
\end{align}
with $Z(p^2)$ the gluon dressing function.

\subsection{Truncation}
In principle one may choose to back-couple the full quark-gluon vertex internally and solve for the 
fully coupled system. This is a straightforward, but highly technical task when real space-like 
momenta are considered. The calculation is significantly increased in complexity when the external 
quark momenta are continued into the complex plane as required in the solution of the quark DSE for 
bound-state studies.

However, this complexity is increased further still when the corresponding quark-antiquark kernel is constructed in accordance with the axial-vector Ward-Takahashi identity~\cite{Munczek:1994zz,Bender:1996bb}. This requires that we evaluate a five-point function that follows from the functional derivative $\dfrac{\delta}{\delta S} \Gamma^\mu_{qqg}$ inherent in the construction of the kernel~\footnote{The situation is simpler in the 3PI formalism, see Ref.~\cite{Sanchis-Alepuz:2015tha}.}. In RL this is trivially vanishing since $\Gamma^\mu_{qqg} \simeq \gamma^\mu$ is not a function of $S$. Beyond this, the five-point function would satisfy a coupled non-linear integral equation~\cite{Bhagwat:2004hn,Matevosyan:2006bk} whose solution is beyond the scope of the present work due to the general kinematics employed here. However, this problem can be made tractable by eliminating the implicit dependence on the quark propagator at some \emph{fixed} loop order. That is, we neglect the explicit back-coupling of the quark-gluon
vertex by replacing the internal quark-gluon vertices of the one-loop Abelian and non-Abelian self-energy diagrams by
\begin{align}
\Gamma^{\mu,\mathrm{internal}}_{qqg}(k,q) = L_1(q^2)\gamma^\mu\;,
\end{align}
where $L_1(q^2)$ is a scalar function of the gluon momentum $q$ that parametrises the missing contributions. The relative quark momentum $k$ is neglected, and 
the asymptotic behaviour $L_1\rightarrow Z_{1f}$ at large momenta ensures multiplicative renormalisability. 
The function $L_1(q^2)$ is constructed iteratively such that it matches the calculated $\gamma^\mu$ component, $h_1$ in Eq.~\eqref{eqn:goodbasis}, inclusive of flavour dependence.

\begin{figure}[!t]
\centering{\resizebox{0.8\columnwidth}{!}{\includegraphics{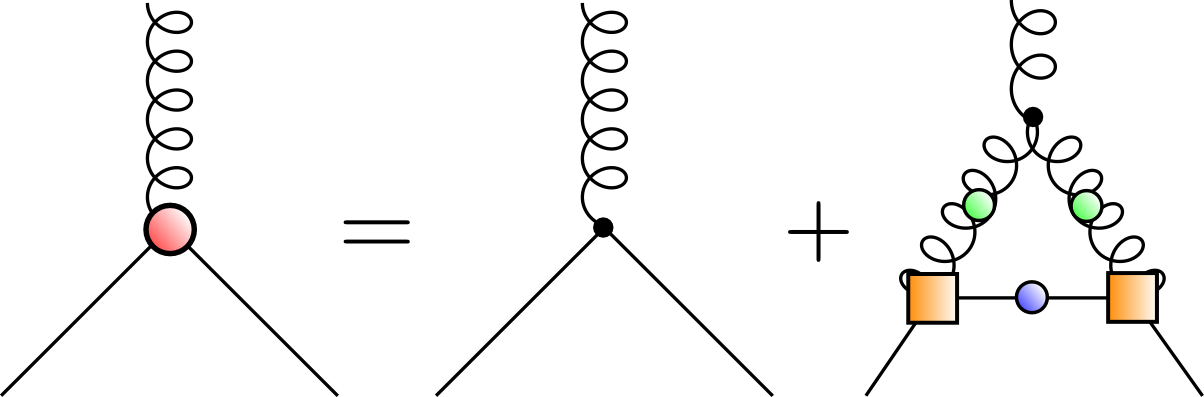}}}
\caption{Truncation of the quark-gluon vertex DSE. Internal quark-gluon vertices (denoted by a square) are provided by Ansatz, see Eq.~\eqref{eqn:intvertex}\label{fig:qgvtrunc}}
\end{figure}
\begin{figure}[!b]
\centering{\resizebox{0.98\columnwidth}{!}{\includegraphics{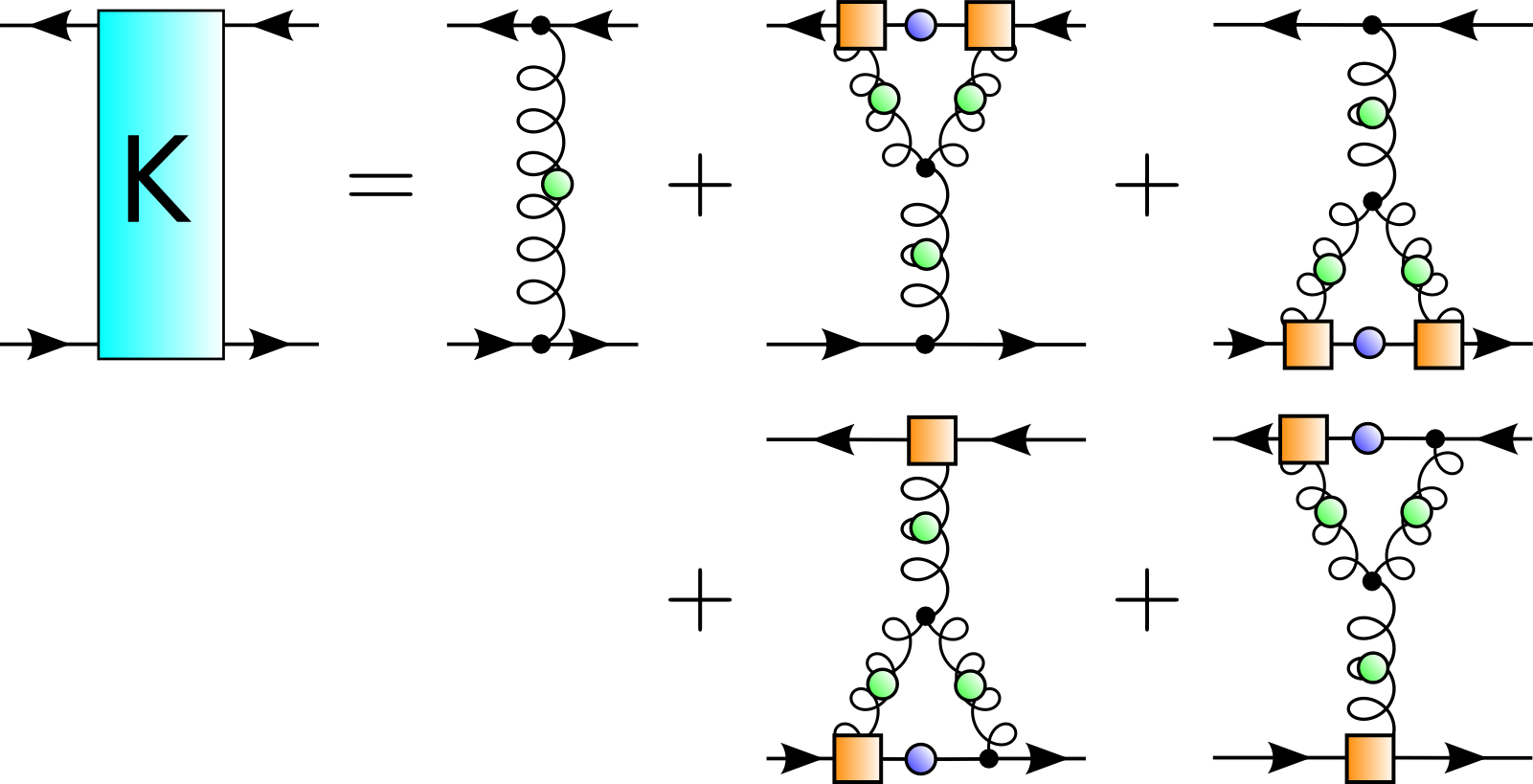}}}
\caption{The corresponding quark-antiquark kernel consistent with chiral symmetry breaking (prefactors omitted).\label{fig:kernelNA}}
\end{figure}
For definiteness, let us take the truncation of the quark-gluon vertex portrayed in Fig.~\ref{fig:qgvtrunc}, which features
no explicit dependence on the quark-gluon vertex itself.  The corresponding chiral symmetry preserving quark-antiquark kernel is given in Fig.~\ref{fig:kernelNA}. The Abelian diagram is neglected for two reasons: firstly, it is subleading by $N_c^2$ with respect to the non-Abelian diagram; secondly, it generates non-ladder like corrections to the bound-state kernel that are significantly more difficult to include~\cite{Munczek:1994zz,Bender:1996bb,Watson:2004kd,Fischer:2009jm,Williams:2009wx}. This will be the focus of future work.

What remains is the specification for the internal dressing of the quark-gluon vertex. In order to be self-consistent,
we chose the following form that reproduces well the $\gamma^\mu$ component, $h_1$, of the numerical calculation
\begin{align}\label{eqn:intvertex}
L_1(p^2) &=h Z_{1f}\bigg\{ \frac{\Lambda(M_0)}{1+y} \nonumber \\
&+\frac{z}{1+z}\left[ \frac{4\pi}{\beta_0 \alpha_\mu} \left( \frac{1}{\log{x}}-\frac{1}{x-1}  \right)  \right]^{18/44} \bigg\}\;.
\end{align}
Here, $h=2.302$, $x = p^2/0.6$, $y = p^2/0.35$, $z = p^2/0.33$, $\beta_0 = 11 N_c/3-2N_f/3$ and $\alpha_\mu=0.7427$. We choose the number of quark flavours $N_f=0$ for consistency with the quenched Yang-Mills sector.  The IR enhancement $\Lambda$ is a 
function of the quark mass at zero Euclidean momenta, $M_0$. Empirically, this is parametrised by the followed rational polynomial
\begin{align}\label{eqn:intvertexlambda}
\Lambda(M_0) = \frac{a + bM_0+cM_0^2}{M_0 + d M_0^2}\;,
\end{align}
with $a \simeq -0.79$, $b \simeq 13.1$, $c \simeq 5.74$, $d \simeq 10.9$. These parameters have been obtained by repeated fitting of the solution until convergence is achieved.

\begin{figure}[!t]
\centering{\resizebox{0.85\columnwidth}{!}{\includegraphics{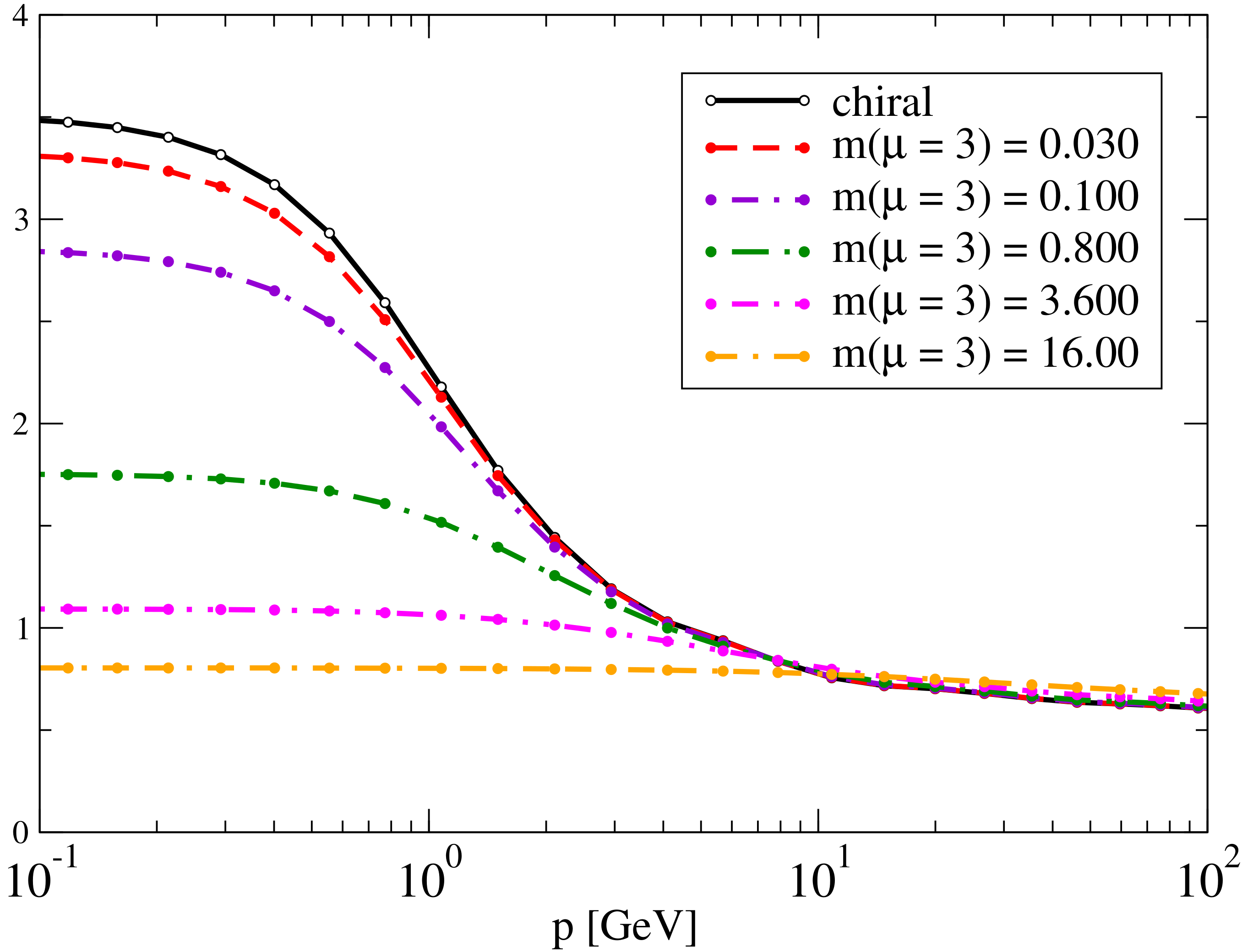}}}
\caption{The scalar dressing of the internal quark-gluon vertex for quarks of increasing mass. For heavy flavours, the vertex tends towards its bare form $Z_{1f}$.\label{fig:veffective}}
\end{figure}

 In Fig.~\ref{fig:veffective} we display the vertex dressing for different quark masses. The important thing to note about this form of the vertex is the implicit flavour dependence. The combination of Eq.~\eqref{eqn:intvertex} and Eq.~\eqref{eqn:intvertexlambda}
is such that for heavy quarks, the vertex dressing reduces to $L_1 \rightarrow Z_{1f}$ but features an enhancement sufficient for DCSB in light quarks.

\section{Results}
We solved the quark and quark-gluon vertex DSE for real space-like momenta according to the method and model outlined above. The calculation is
quenched in the sense that quark-loop contributions to the gluon propagator are not considered. However, since the flavour dependence and unquenching effects present in a dressed quark-gluon vertex are generally of more interest here, see Ref.~\cite{Fischer:2005en} vs. 
Ref.~\cite{Fischer:2007ze,Fischer:2008wy}, this proves to be a good approximation. Unquenching effects in the Yang-Mills sector can be easily accommodated by modification of the ghost and gluon propagators.

With a bare three-gluon vertex, the overall renormalisation constant of the non-Abelian diagram is $Z_1$. Renormalisation of the quark-gluon vertex DSE itself is provided by the inhomogeneous term $Z_{1f}\gamma^\mu$. Note that these are not independent constants and are related to other renormalisation constants through Slavnov-Taylor identities,
\begin{align}
Z_1 = \frac{Z_3}{\tilde{Z}_3}\;,\;\;\; Z_{1f} = \frac{Z_2}{\tilde{Z}_3}\;
\end{align}

We determined the chiral condensate from the trace of the quark propagator. Converting to $\overline{MS}$ at $\mu=2$~GeV we found $-\left(\langle \bar{\psi}\psi\rangle\right)^{1/3}=275$~MeV in good agreement with other phenomenological studies.

For the purposes of presentation, it is more enlightening to rotate the transverse orthonormal basis into one that 
exhibits charge conjugation invariance~\cite{Eichmann:2014qva}. Specifically, we use the variation~\footnote{We thank Gernot Eichmann for highlighting the efficacy of this basis.}
 \begin{align}\label{eqn:goodbasis}
T^{\mu\nu}_{(k)} \Gamma^\nu(l,k) &= \sum_{i=1}^8 h_i X_i^\nu(l,k) \nonumber\\
						     &=h_1\, \gamma^\mu_T 
                             + h_2\, l^\mu_T \slashed{l}
                             + h_3\, i l^\mu_T 
						     \nonumber\\
                            &+ h_4\, \left(l\cdot k\right) \frac{i}{2}\left[\gamma^\mu_T,\slashed{l}\right]
                             + h_5\, \frac{i}{2}\left[\gamma^\mu,\slashed{k}\right]
                             \nonumber\\
                            &+ h_6\, \frac{1}{6}\left[\gamma^\mu,\slashed{l},\slashed{k}\right]
                             + h_7\, t^{\mu\nu}_{(kl)}\left(l\cdot k\right) \gamma^\nu
                             \nonumber\\
                            &+ h_8\, t^{\mu\nu}_{(kl)}\frac{i}{2}\left[\gamma^\nu,\slashed{l} \right]\,,
\end{align}
which defines the basis elements $X_i^\nu$. Here, $\left[A,B,C\right]=\left[A,B\right]C+\left[B,C\right]A+\left[C,A\right]B$, the incoming gluon momentum
is $k^\mu=p_3^\mu$, and $l^\mu=\left(p_1^\mu+p_2^\mu\right)/2$ is the relative quark momentum. Quantities with
a subscript $T$ are contracted with the transverse projector $T^{\mu\nu}_{(k)}$, see Eq.~\eqref{eqn:gluonprop}. Additionally,
$\tau^{\mu\nu}_{(kl)}=\left(k\cdot l\right)\delta^{\mu\nu}-l^\mu k^\nu$.

\begin{figure}[!thp]
\centering{
		   \resizebox{0.84\columnwidth}{!}{\includegraphics{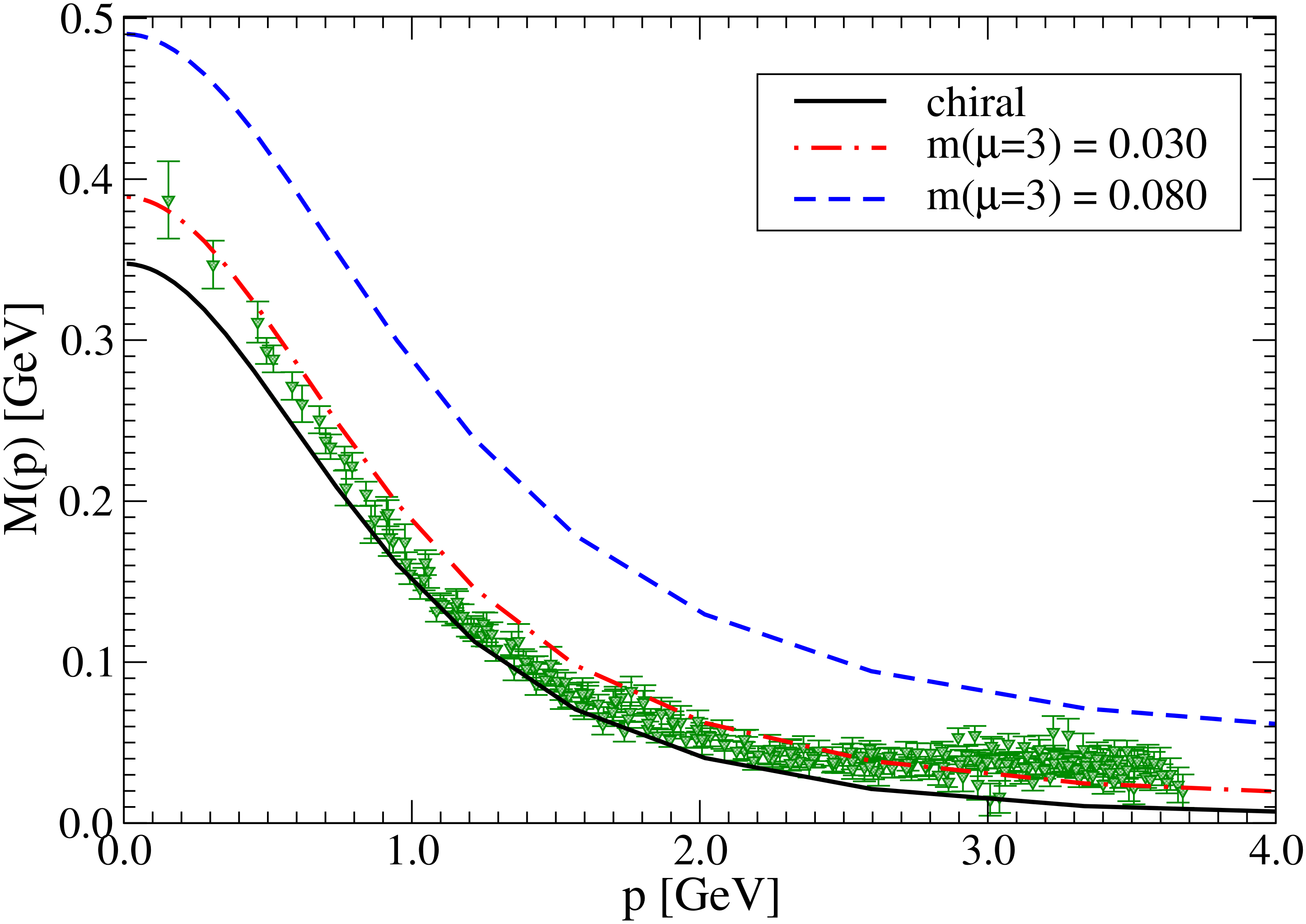}}	   
           \resizebox{0.84\columnwidth}{!}{\includegraphics{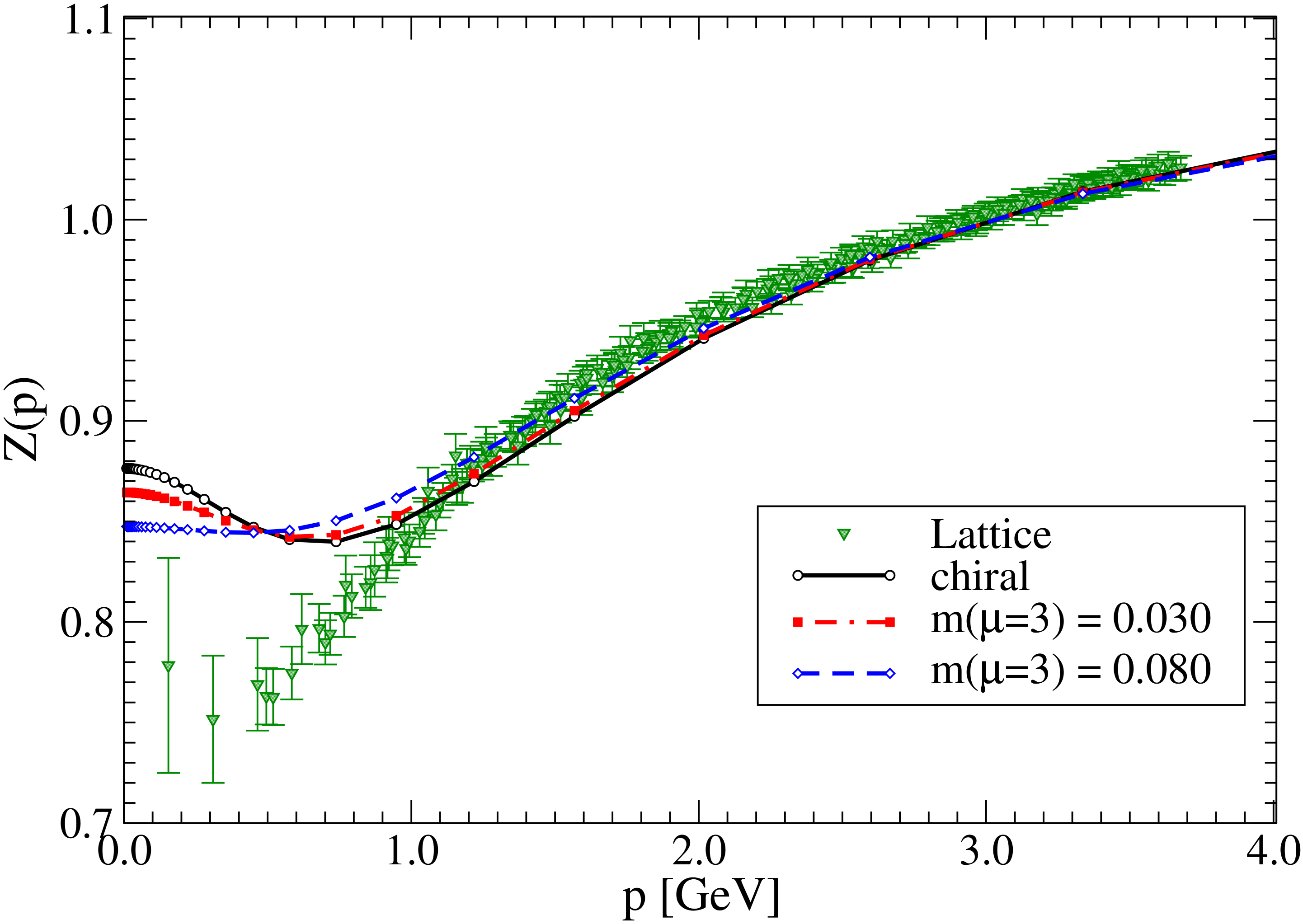}}
           }
\caption{The quark mass function $M(p)$ (top) and quark wavefunction (bottom) $Z(p)$ for different values of the quark mass in the infrared, with comparison to the lattice data of Bowman et al. \cite{Bowman:2005vx}. Dimensionful quantities are in GeV.}\label{fig:resultsZM}
\end{figure}

\subsection{Flavour dependence}

In Fig.~\ref{fig:resultsZM} we show the quark mass function and quark wavefunction for chiral, $u/d$, $s$, $c$ and $b$ quarks. These are further compared, in the case of light quarks, to a quenched lattice calculation wherein a similar qualitative behaviour is observed. The resulting quark mass function and quark wavefunction have the same behaviour as typically seen in other Dyson--Schwinger calculations, which reassuringly confirms that even simple models can perform adequately in this regard.

\begin{figure}[!thp]
\centering{\resizebox{0.99\columnwidth}{!}{\includegraphics{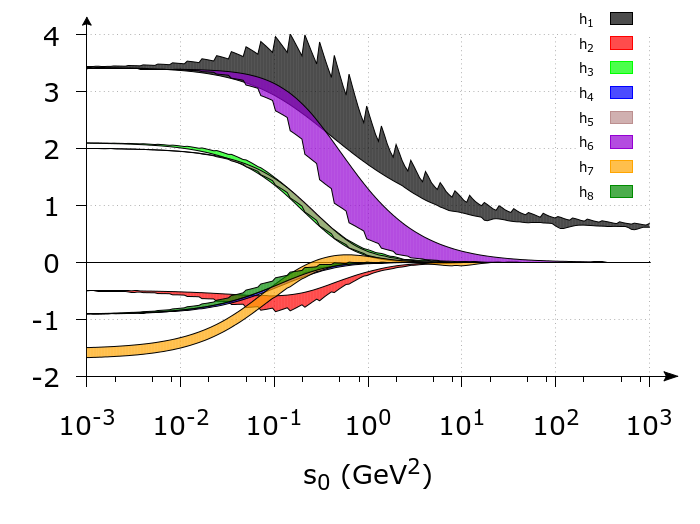}}
           \resizebox{0.99\columnwidth}{!}{\includegraphics{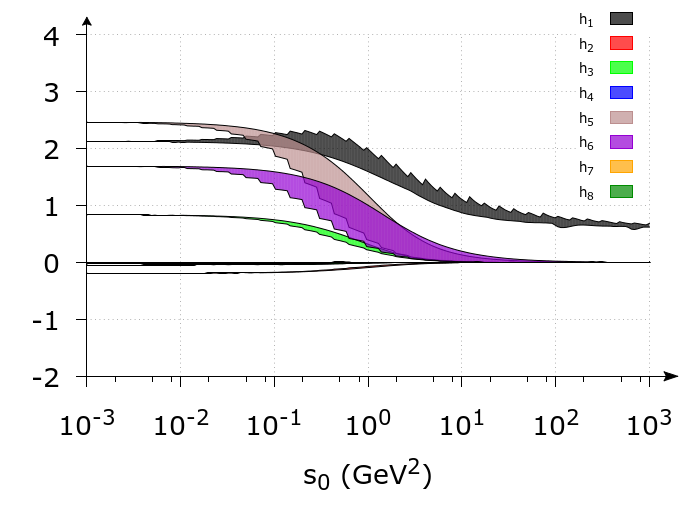}}
           \resizebox{0.99\columnwidth}{!}{\includegraphics{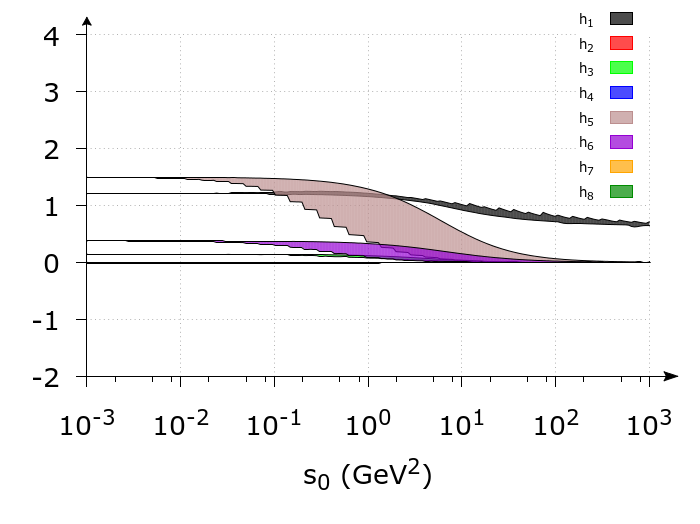}}
}
\caption{(colour online) The dressing functions $h_i$ of the quark-gluon vertex in the charge-conjugate symmetric basis of Eq.~\eqref{eqn:goodbasis} for:
(top) up/down quark $m(\mu)=0.03$~GeV; (middle) charm quark $m(\mu)=0.8$~GeV; (bottom) quark $m(\mu)=3.6$~GeV. We use $s_0=l^2/3+k^2/4$ as a symmetric variable with bands corresponding to the angular dependence, and renormalisation point $\mu=3$~GeV.}\label{fig:resultsVD}
\end{figure}

Since the calculation of the quark-gluon vertex depends both implicitly and explicitly on the quark flavour, we expect to see this reflected strongly in the vertex dressings. We combine the relative quark momentum $l$ and incoming gluon momentum $k$ into the variables
\begin{align}
s_0 =\frac{l^2}{3}+\frac{k^2}{4}\;,\;\;
a   =\frac{l\cdot k}{\sqrt{3}s_0}\;,\;\;
s   =1-\frac{k^2}{2s_0}\;.
\end{align}
Here $s_0$ is proportional to the sum of the incoming squared momenta and $a,s \in \left[-1,1\right]$ are two angular variables.
In Fig.~\ref{fig:resultsVD} we plot the dressing functions $h_1\ldots h_8$ as a function of $s_0$, with the shaded band corresponding to the
combined dependence on $a,s$ (spikes are a plotting artefact). 
As expected, in this transverse basis the dominant component for all quark flavours is the tree-level term $\gamma^\mu_T$ which tends towards $Z_{1f}$ logarithmically for large space-like momenta. It is immediately obvious from the size of the shaded regions that the angular dependence is relatively sizeable
for the majority of basis coefficients. For light quarks, the next largest components are $h_6$, followed by $h_5$. These are the vector and scalar components of the anomalous chromomagnetic moment, whose relevance to the $a_1$ and $\rho$ splitting has been highlighted in Refs.~\cite{Chang:2011ei,Cloet:2013jya}.

As the quark mass is increased (middle and lower graph), $h_5$ is less suppressed with regards to $h_6$ owing to its mass dimensions. That $h_5$ remains sizeable for the bottom quark may be significant for the spectrum of excited states, since rainbow-ladder (essentially the $h_1$ component alone) already fares well for ground-states of bottomonium~\cite{Blank:2011ha}; similar expectations apply to charmonium.   These two quantities are certainly of interest to hadron phenomenology in a study of bound-states beyond rainbow-ladder.

\begin{figure}[!t]
\centering{\resizebox{0.95\columnwidth}{!}{\includegraphics{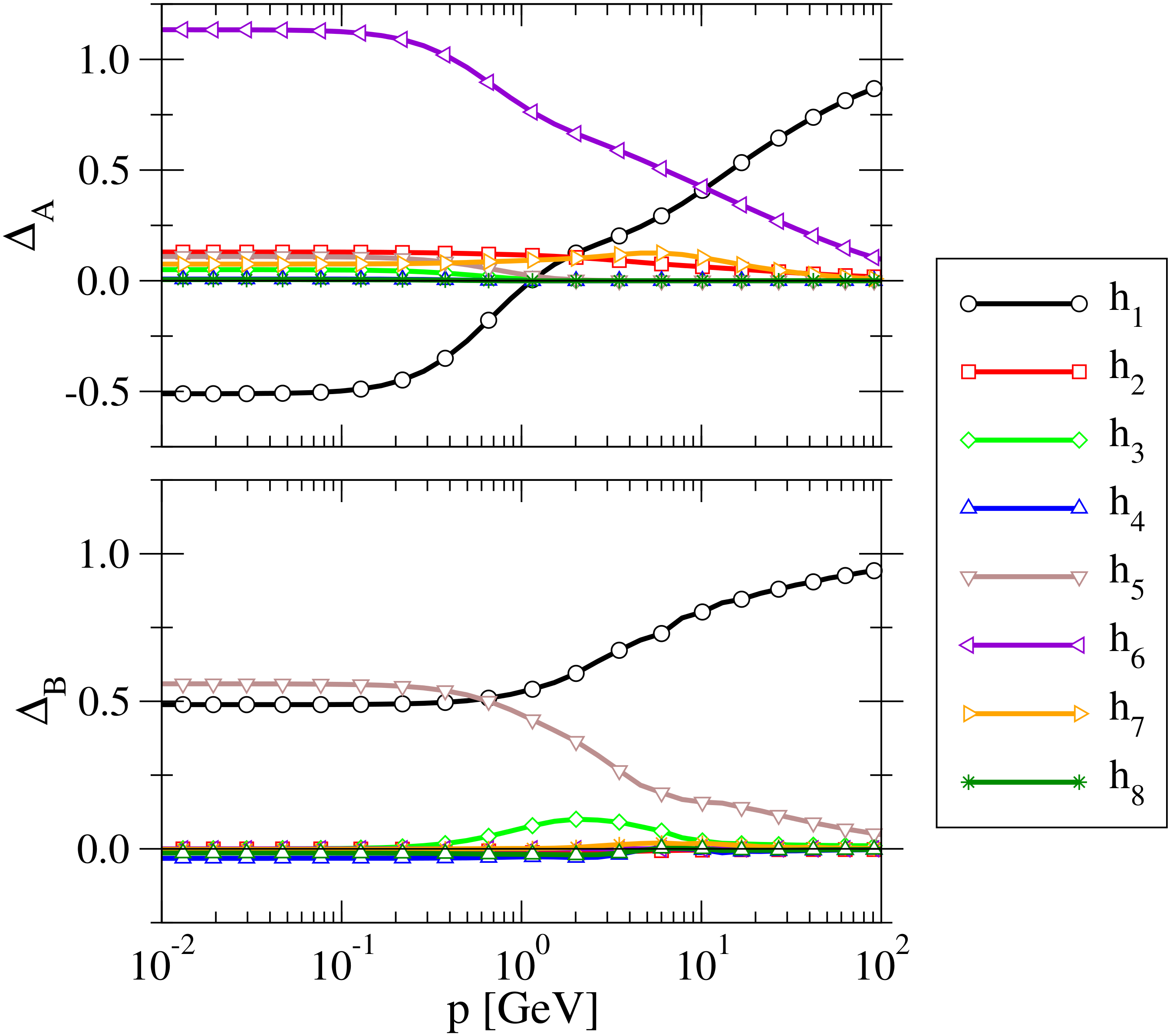}}}
\centering{\resizebox{0.95\columnwidth}{!}{\includegraphics{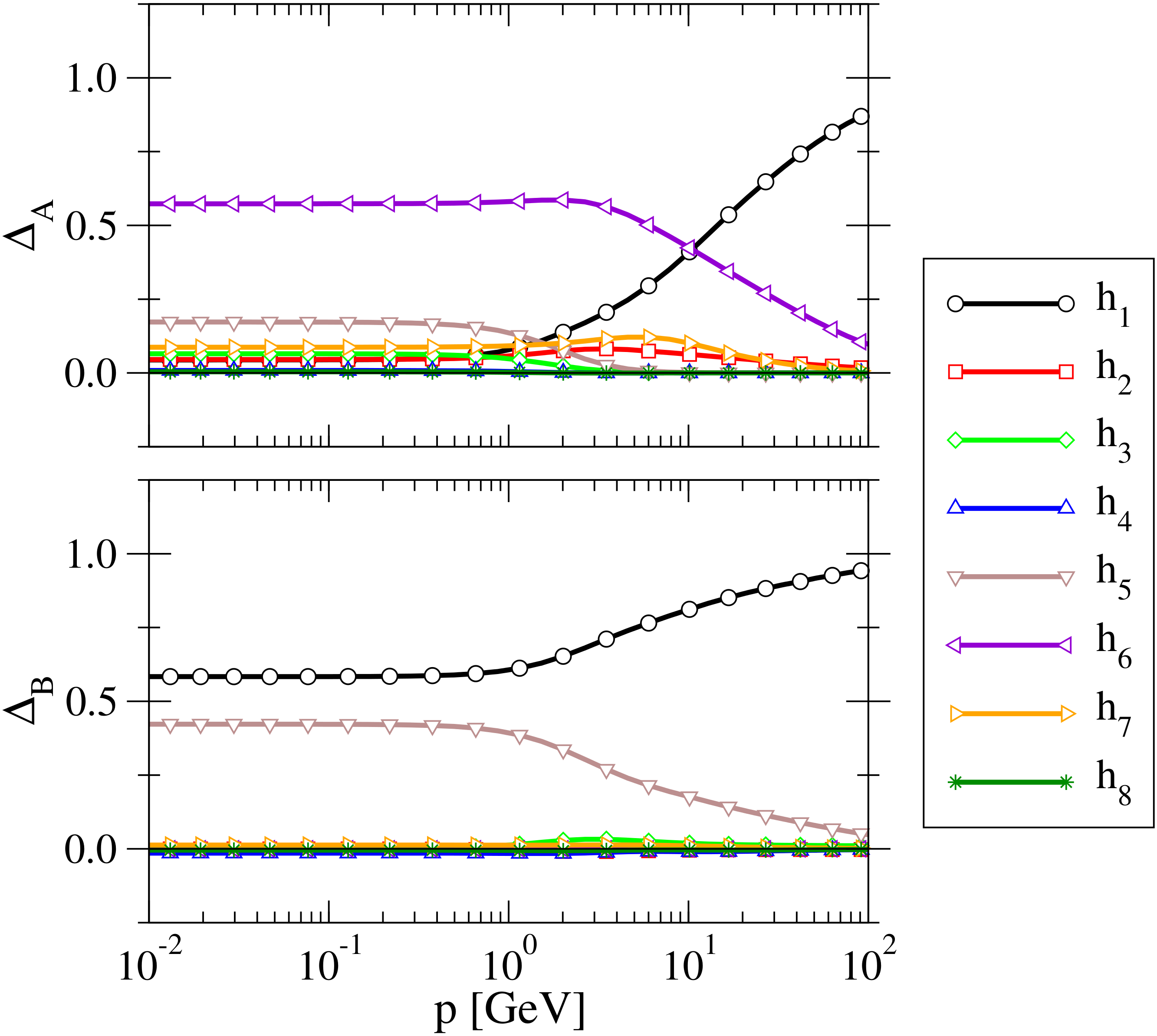}}}
\caption{(colour online) Relative contribution of the vertex components, Eq.~\eqref{eqn:goodbasis}, to the vector ($\Delta_A$) and scalar part ($\Delta_B$) of the quark propagator, Eq.~\eqref{eqn:selfratio} for: (left) a light quark of mass $0.03$~GeV at the renormalisation point $\mu=3$~GeV; (right) a heavy quark of mass $0.8$~GeV at the renormalisation point $\mu=3$~GeV. The contribution $h_6$ to the scalar part of the propagator is always zero.}\label{fig:resultsweight}
\end{figure}

Comparing light to heavy quarks (top vs. middle and lower graphs), we see that the $h_2$, $h_3$ and $h_5$ components are significant in the former, being suppressed for heavier masses. These will contribute towards the fine details of the mass splitting between states, which are lacking for many ground
and excited state mesons as can be seen in e.g. Ref.~\cite{Fischer:2014xha}.

However, to judge the relevance of each vertex component we have to investigate its contribution to some quantity. For example, one could take the chiral condensate, but this would only consider the Dirac even parts of the vertex. A better test is to investigate the weighted contribution of each component $h_i$ to the self-energy of the quark DSE under projection onto the vector and scalar parts of the propagator, Fig.~\ref{fig:resultsweight}.   Following Eq.~\eqref{eqn:quarkdse} and Eq.~\eqref{eqn:goodbasis} we define
\begin{align}\label{eqn:partofself}
  \Sigma^i_A(p) &= \Tr\left[\slashed{p}\int d^4k\gamma_\mu S(p-k) \left[h_i X^i_\nu(l,k)\right]\right]D_{\mu\nu}(k)\;,\nonumber\\
  \Sigma^i_B(p) &= \Tr\left[\ONE\int d^4k\gamma_\mu S(p-k) \left[h_i X^i_\nu(l,k)\right]\right]D_{\mu\nu}(k)\;,
\end{align}
from which we construct the relative integrated contribution of the vertex components to the vector part by
\begin{align}\label{eqn:selfratio}
\Delta^i_A(p) = \frac{\Sigma^i_A(p)}{\sum_{i=1}^8 \Sigma^i_A(p)}\;,
\end{align}
and similarly for the scalar part $\Delta^i_B(p)$.
This enables the kernel function to be taken into account since it can have a dramatic influence in terms of enhancement or suppression.  Summing up all of the curves gives the total of one, which would be the result for rainbow-ladder (or any theory with just one basis component). We see that, as expected, the $h_1$ component is dominant 
at large momenta since the remaining coefficients $h_i$ fall off with $p^2$. We clearly see the importance of the anomalous chromomagnetic moment terms $h_6$ for the vector projection (top line) and $h_5$ for the scalar projection (bottom line). The scalar part of the vertex $h_3$, traditionally thought of as being important with regards to communicating effects of dynamical chiral symmetry breaking, is much less relevant for both the scalar and vector projections. These
statements are essentially independent for reasonable ranges of quark masses.

\subsection{Running coupling}

\begin{figure}[!t]
\centering{
\resizebox{0.85\columnwidth}{!}{\includegraphics{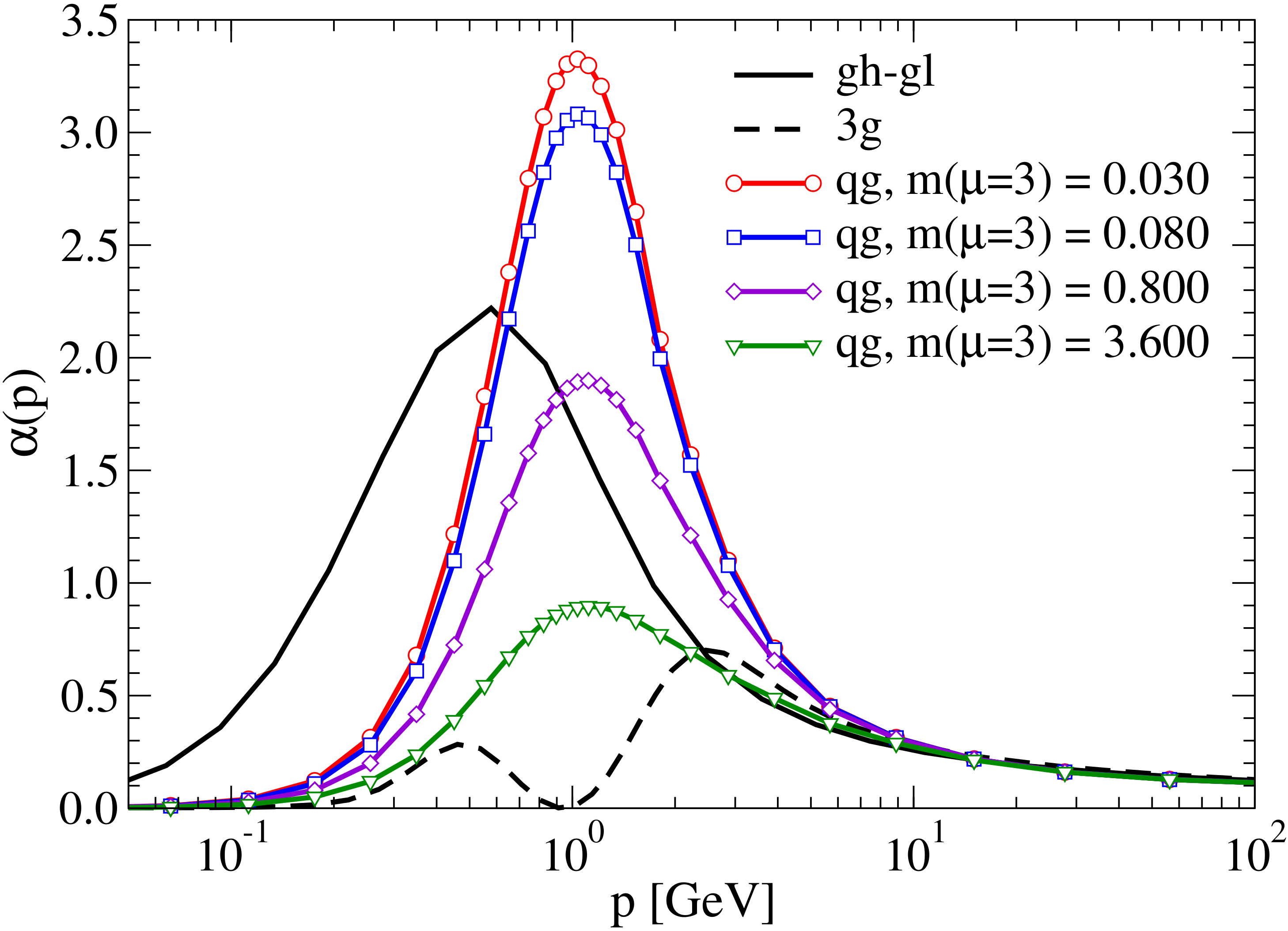}}}
\caption{The running coupling of the quark-gluon vertex, for different quark flavours, compared to the running of the ghost-gluon vertex and the three-gluon vertex. Dimensionful units are in GeV.\label{fig:running}}
\end{figure}

The running couplings from the ghost-gluon, three-gluon and quark-gluon vertices can be defined
\begin{align}
\alpha_\text{gh-gl} =\frac{g^2}{4\pi}\, Z\,G^2\,,\;\;\;
\alpha_\text{3g} =\frac{g^2}{4\pi}\, Z^3\,\Gamma_\text{3g}\,,  \\
\alpha_\text{qg} =\frac{g^2}{4\pi}\, Z\,\left(Z_f\,\Gamma_\text{qg}\right)^2\,.\qquad
\end{align}
Here, $\Gamma_\text{gh}$, $\Gamma_\text{3g}$ and $\Gamma_\text{qg}$ denote the tree-level dressing functions of the ghost-gluon, three-gluon and quark-gluon vertex, respectively, and $Z$, $G$ and $Z_f$ are the dressings in the gluon, ghost and quark propagators.  They depend on a momentum scale $\mu$, which for the three-point functions we define via the asymmetric momentum routing $p^2=p_1^2=2p_2^2=3p_3^2$. We display these running couplings as a function of this momentum scale in Fig.~\ref{fig:running}, where the running of the quark-gluon vertex is given for several different quark flavours.
As expected, the different couplings deviate for non-perturbative scales but agree well in the UV; this agreement can be improved by including a
dressed three-gluon vertex or RG improvment term in the non-Abelian diagram.

\subsection{Quark for complex momenta}
\begin{figure}[!ht]
\centering{
\resizebox{0.73\columnwidth}{!}{\includegraphics{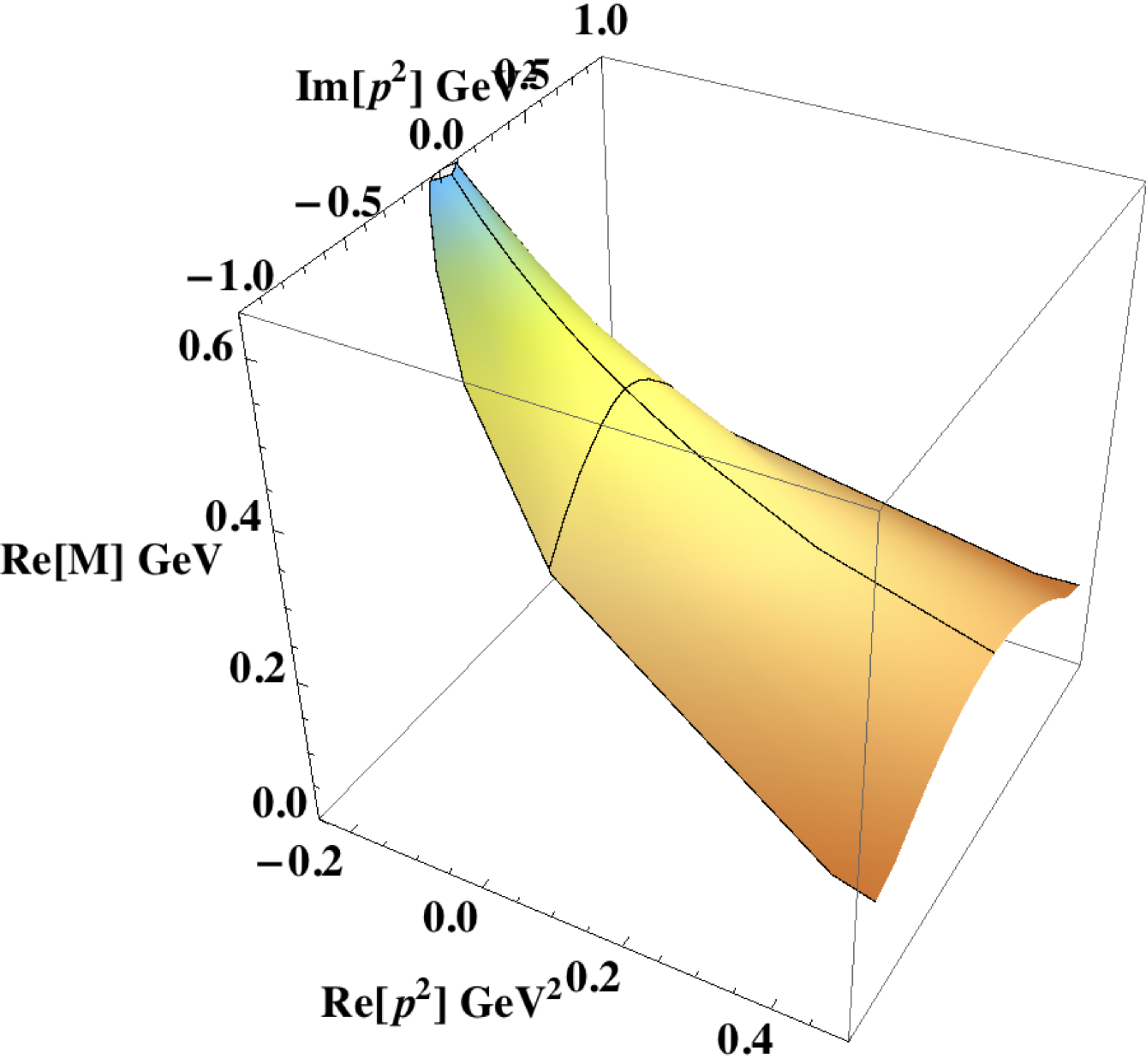}}
\resizebox{0.73\columnwidth}{!}{\includegraphics{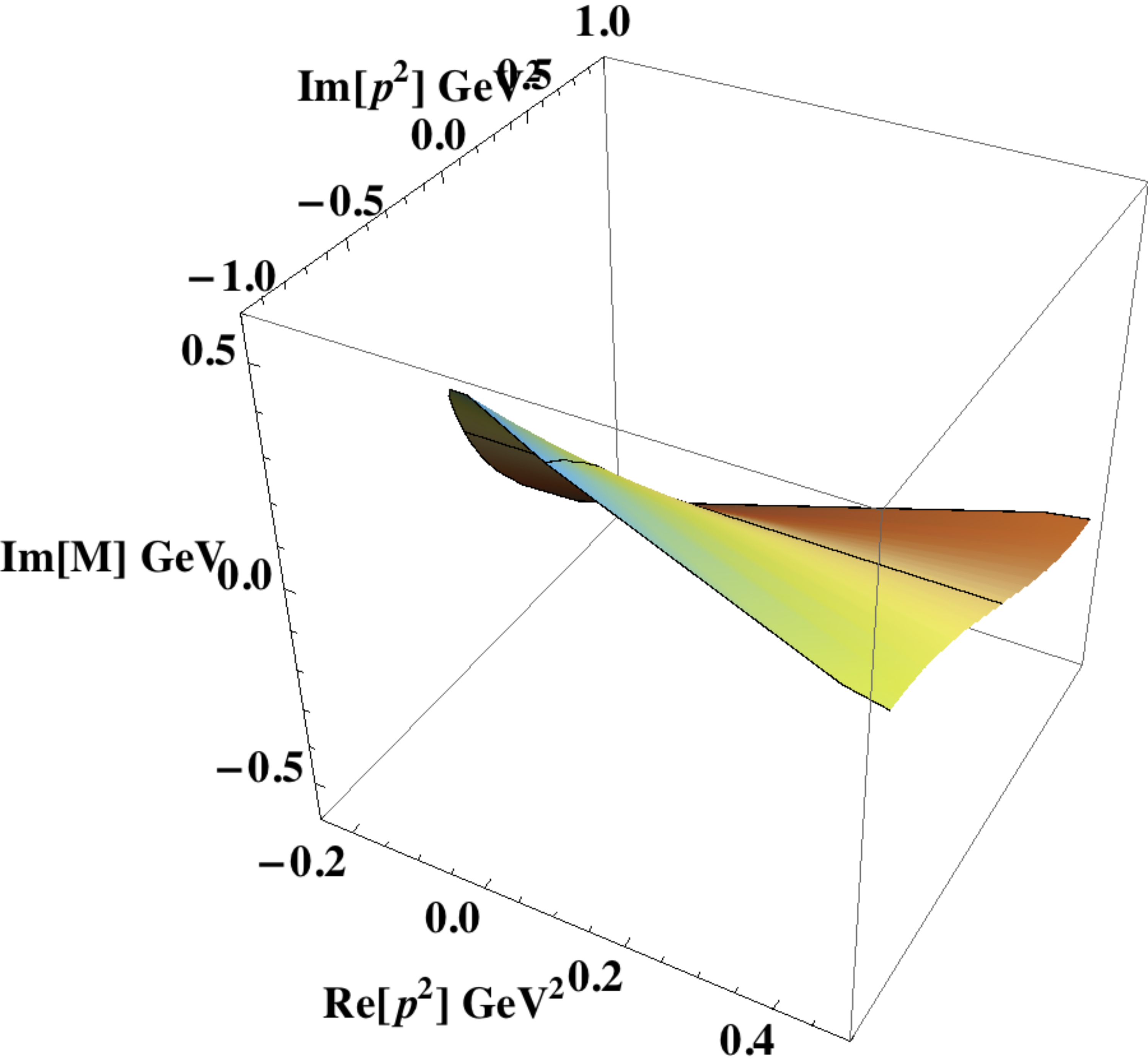}}}
\caption{The real and imaginary parts of the quark dressing function $M$ in the complex-plane.
\label{fig:cmplxsolM}}
\end{figure}

In Fig.~\ref{fig:cmplxsolM} we give the real and imaginary parts of the quark mass function of light $u/d$ quarks in a region of the complex plane centred on the origin; those for the quark wavefunction $Z_f$, are similarly smooth. They are obtained via the shell-method as outlined in the introduction, with the quark-gluon vertex similarly analytically continued to the complex plane. Note that the vertex is similarly analytic in the region of the complex plane considered.

Each macro cycle in which the quark-gluon vertex is updated and the quark propagator solved, takes approximately three minutes on a modest single CPU core and is easily parallelised. Introducing a fully dressed three-gluon vertex does not impact upon performance appreciably. The introduction of fully dressed internal quark-gluon vertices will scale the algorithmic difficulty by approximately eight for each vertex, plus additional overhead for the interpolation of the dressing functions. Compared to this macro cycle, the micro cycle in which the quark-propagator is solved is essentially for free. The process is iterated until convergence is reached.

With the quark propagator thus obtained in the complex plane, we may proceed to construct the quark-gluon vertex as relevant for bound-state studies, using the enlarged basis given in Appendix~\ref{sec:bsebasis} with a greater number of kinematic invariants. However, this procedure requires no further iteration and so results may be pre-calculated and tabulated for later use.

\section{Conclusions and Outlook}
We presented an adaptable approach to the decomposition and calculation of the quark-gluon vertex for both real and complex Euclidean momenta, as required for studies of hadronic bound-states. A suitable model was presented that features dynamical chiral symmetry breaking, producing quark propagators that are compatible with lattice results. The quark-flavour dependence and the effective quark-gluon vertex were found to be sizeable, reducing to the expected single one-gluon exchange in the heavy quark limit. However, in contrast to previous expectations it is the anomalous chromomagnetic moment contributions, arising from dynamical breaking of chiral symmetry that dominate over traditional scalar contributions. This is certainly of relevance to hadronic bound-states, currently under investigation for both mesons and baryons~\cite{SanchisAlepuzWilliams}, and highlights the inadequacy of simple RL studies of baryons and mesons which typically feature a flavour independent vector-like interaction.
      
There are several improvements that can be made. Firstly, whilst preserving the connection to a chiral symmetry preserving truncation of the quark-(anti)quark interaction, one could include an explicit solution of the three-gluon vertex in the non-Abelian contribution to the quark-gluon vertex. However, there one would need to include quark-loop effects and perhaps account for two-loop contributions in order for the result to be reliable. Secondly, one may include explicitly the Abelian contribution. This is a simple task for the quark-gluon vertex DSE, but introduces crossed-ladder kernels in the BSE kernel without introducing sizeable corrections. Thirdly, complicating the preservation of the axWTI and requiring solution of a coupled five-point function, one may choose to back-couple the quark-gluon vertex internally as considered in Ref.~\cite{Hopfer:2013np}. Finally, one may adjust the internal modelling of the internal vertices so that the final result is in agreement with alternative diagrammatic resummations such as the functional renormalisation group~\cite{Mitter:2014wpa}, and future Lattice investigations of this object.

\section*{Acknowledgments}
RW would like to thank R.~Alkofer, G.~Eichmann, C.~S.~Fischer, M.~Mitter, H.~Sanchis-Alepuz and M.~Vujinovic for useful discussions and a critical reading of the manuscript. 
This work was supported by the Helmholtz International Center for FAIR within the LOEWE
program of the State of Hesse, 
by BMBF under contract 06GI7121, and the Austrian Science Fund (FWF) under project number M1333-N16.

\appendix
\numberwithin{equation}{section}

\section{Vertex basis for quark DSE}

\subsection{Transverse}\label{sec:dsetrans}

In the case of real Euclidean momenta, the description of the quark-gluon vertex is
straightforward. Given that we have two Dirac indices, one Lorentz index and two
independent momenta $p_1$, $p_2$ the most naive basis decomposition would be
\begin{align}
\left(\begin{array}{ccc}
\gamma^\mu, & p_1^\mu, & p_2^\mu \\
\end{array}\right)
\times
\left(\begin{array}{cccc}
\ONE, & \slashed{p}_1, & \slashed{p}_2, & \slashed{p}_1\slashed{p}_2 \\
\end{array}\right)\;,
\end{align}  
which features $12$ components. This is of course not unique as we can construct 
different linear combinations of these basis elements, for instance such that the
vertex is free of kinematic singularities~\cite{Ball:1980ay,Kizilersu:1995iz}. 

We specialise to the case of DSEs in Landau gauge, wherein 
every Lorentz index will ultimately be 
contracted by the transverse projector $T^{(k)}_{\mu\nu} = \delta_{\mu\nu}-k_\mu k_\nu/k^2$ 
contained within the gluon propagator. If 
we define the total incoming momentum of the gluon $k^\mu = p_2^\mu-p_1^\mu$
and the relative quark momentum\footnote{Note that in practical calculations it is more prudent to employ
unequal momentum partitioning in the quark-gluon vertex when defining the relative momentum $l$.} $l^\mu = (p_2^\mu+ p_1^\mu)/2$, we can
construct the following orthonormal elements
\begin{align}
t^\mu &= \hat{k}^\mu\,, \\
s^\mu &= \widehat{T^{(t)}_{\mu\nu}l^\nu}\,, \\
\gamma_{TT}^\mu &:=T^{(t)}_{\mu\alpha}T^{(s)}_{\alpha\nu}\gamma^\nu = \gamma^\mu -\slashed{t}\,t^\mu- \slashed{s}\,s^\mu\;.
\end{align}
where the hat indicates normalisation. These 
then provide the orthogonal basis
\begin{align}
\left(\begin{array}{ccc}
\cellcolor{blue!20}\gamma^\mu_{TT}, & \cellcolor{blue!20}s^\mu, & t^\mu \\
\end{array}\right)
\times
\left( \ONE,\;\; {\slashed{s}},\;\; {\slashed{t}},\;\; {\slashed{s}}\,{\slashed{t}}\,\right)
\;,
\end{align}
where $s\cdot t=0$, and $\gamma^{TT}_\mu s^\mu = \gamma^{TT}_\mu t^\mu = 0$. 
For later convenience we
denote this Dirac part by
\begin{align}
R^{i}_{(\ONE)} =
\left( \ONE,\;\; {\slashed{s}},\;\; {\slashed{t}},\;\; {\slashed{s}}\,{\slashed{t}}\,\right)\;,
\end{align}
Due
to the transversal nature of Landau gauge only those highlighted components contribute, thus requiring just
$8$ components to completely describe the quark-gluon vertex.  Being orthogonal, the projectors for the scalar
coefficients are easy to construct.

Let us focus on a particular choice of momentum frame. For squared momenta $k^2$, $l^2$ and their cosine $z=\widehat{k\cdot l}$ we can write
\begin{align}
k^\mu &= \left|k\right|\, (\begin{array}{cccc} 0, & 0, & 0, & 1 \end{array})\;, \\
l^\mu &= \left|l\right|\, (\begin{array}{cccc} 0, & 0, & z^\prime, & z \end{array})\;,
\end{align}
where $z^\prime = \sqrt{1-z^2}$. The orthonormal momenta, specifically for this frame, reduce to
\begin{align}
t^\mu = ( \begin{array}{cccc} 0, & 0, & 0, & 1 \end{array}) \;,\nonumber\\
s^\mu = ( \begin{array}{cccc} 0, & 0, & 1, & 0 \end{array}) \;,\nonumber\\
r^\mu = ( \begin{array}{cccc} 0, & 1, & 0, & 0 \end{array}) \;,\nonumber\\
v^\mu = ( \begin{array}{cccc} 1, & 0, & 0, & 0 \end{array}) \;.\label{eqn:ortho}
\end{align}
We introduced the orthonormal vectors $r^\mu$, $v^\mu$ to completely
span the vector space. 
It is clear that we may then write
\begin{align}
\gamma^\mu_{TT} = \slashed{v}\,v^\mu + \slashed{r}\,r^\mu\;,
\end{align}
which allows us to separate the Dirac from the Lorentz parts
\begin{align}
L_{(H)}^\mu &= H^\mu\,,\,\;\; \qquad\;\textrm{for~}H=\left\{ v,r,s,t \right\} \,,\label{eqn:LT1}\\
R^{i}_{(H)} &= \left\{ 
	\begin{array}{ ll}
		\phantom{\slashed{H}}R^{i}_{(\ONE)}
			& \quad \textrm{for~}H =\left\{s,t\right\} \\
		\slashed{H}R^{i}_{(\ONE)}
			& \quad \textrm{for~}H =\left\{r,v\right\} 
	\end{array}
	\right.\label{eqn:LT2}
\end{align}
The basis is then given by
\begin{align}
\begin{array}{ccc}
\cellcolor{blue!20}L_{(v)}^\mu R^i_{(v)} + L_{(r)}^\mu R^i_{(r)}\,,&\cellcolor{blue!20} L_{(s)}^\mu R^i_{(s)}\,,& L_{(t)}^\mu R^i_{(t)}\;,
\end{array}
\end{align}
where once again only those elements relevant in Landau gauge are highlighted. The reduced
quark-gluon vertex, as a function of the relative and total momentum $l$ and $k$, is then written
as
\begin{align}
\Gamma^\mu(l,k) = \sum_{i=1}^4\sum_{a=\{v,r,s\}}c_{a}^{i}(l,k) L^\mu_{(a)}R_{(a)}^i\;,
\end{align}
and the projectors, defined $\bar{R}^i$ and $\bar{L}_a^\mu$, satisfy
\begin{align}
\Tr\left[ \bar{R}^i_{(a)} R^{i^\prime}_{(a^\prime)}\right] &= \delta_{aa^\prime}\delta_{ii^\prime}\,,\qquad
 \bar{L}^\mu_{(a)} L^{\mu}_{(a^\prime)} &= \delta_{aa^\prime}\,,
\end{align}
such that
\begin{align}\label{eqn:basisprojection1}
c_{(a)}^{i}(l,k) = \bar{L}^\mu_{(a)} \Tr\left[ \bar{R}^i_{(a)} \Gamma^\mu(l,k)\right]\;.
\end{align}
No summation over $a$ is implied on the right-hand side.
In practical calculations, we reconstruct the $\gamma^\mu_{TT}$
\begin{equation}
c^{i}_{(v)} L^\mu_{(v)} R^i_{(v)} + c^{i}_{(r)} L^\mu_{(r)} R^i_{(r)} = c^i_{(\gamma_{TT}^\mu)} \gamma^\mu_{TT} R^{i}_{(\ONE)}\,,
\end{equation}
to reduce the number of terms. This is useful, but not necessary,
when the back-coupling of vertices in the DSE is required.

\subsection{Non-transverse}\label{sec:dsenontrans}

Instead of the total/relative momentum basis above, which employs the transverse nature
of Landau gauge to eliminate the $L_{(t)}^\mu R^i_{(t)}$ basis elements defined in Eqs.~(\ref{eqn:LT1}--\ref{eqn:LT2}), we
can use a different set of orthogonal but non-transverse basis elements
\begin{align}
t^\mu &= \hat{p}_2^\mu\,, \\
s^\mu &= \widehat{T^{(p_2)}_{\mu\nu}p_1^\nu}\,, \\
\gamma_{TT}^\mu &:=T^{(t)}_{\mu\alpha}T^{(s)}_{\alpha\nu}\gamma^\nu = \gamma^\mu -\slashed{t}\,t^\mu- \slashed{s}\,s^\mu\;.
\end{align}
where once again the hat indicates normalisation. A particular choice of momentum frame is
\begin{align}
p_1^\mu &= \left|p_1\right|\, (\begin{array}{cccc} 0, & 0, & z^\prime, & z \end{array})\;, \\
p_2^\mu &= \left|p_2\right|\, (\begin{array}{cccc} 0, & 0, & 0, & 1 \end{array})\;,
\end{align}
with $z = \widehat{p_1\cdot p_2}$ and $z^\prime = \sqrt{1-z^2}$. Then Eqs.~(\ref{eqn:ortho})--(\ref{eqn:basisprojection1}) follow, with 
the only difference that the $a=\left\{t\right\}$ element is also included in the summation.

\section{Vertex basis for meson BSE}\label{sec:bsebasis}
Due to the kinematics of the meson BSE, the determination of the quark-gluon vertex requires further analytic continuation of the momentum variables.

Referring to Fig.~\ref{fig:qgv-momenta}, we define the external momenta $p_i$ in terms of the
total incoming momentum of the gluon, $\Delta^\mu$, and two relative momenta $\Sigma^\mu$ and $\Omega^\mu$
\begin{align}
p_1^\mu &= \left(\Sigma+\Omega\right)^\mu\;, \nonumber\\
p_2^\mu &= \left(\Sigma+\Omega\right)^\mu+\Delta^\mu\;, \\
p_3^\mu &= \Delta^\mu\;.\nonumber
\end{align}
Here, $\Sigma$ and $\Delta$ are real Euclidean momenta whereas $\Omega$ contains the
complex total momentum of the meson.

In bound-state calculations, a convenient explicit realisation of these momenta $\Delta$, $\Sigma$ and $\Omega$ is
\begin{align}
\Delta^\mu &= \left|\Delta\right| \left(\begin{array}{cccc} 0, & w^\prime y^\prime, & w y^\prime, & y\end{array}\right)\,, \\
\Sigma^\mu &=  \left|\Sigma\right|\,  \left(\begin{array}{cccc} 0, & 0, & z^\prime, & z\end{array}\right)\,, \\
\Omega^\mu &=  \left|\Omega\right|\,  \left(\begin{array}{cccc} 0, & 0, & 0, & 1\end{array}\right)\,,
\end{align}
where $w^\prime = \sqrt{1-w^2}$, $y^\prime = \sqrt{1-y^2}$, $z^\prime = \sqrt{1-z^2}$.
That the vectors $\Sigma$ and $\Omega$ are not parallel, as in the case of the quark DSE, is the cause of the
additional analytic continuation needed for the vertex in the BSE. To accommodate this additional
angular dependence, it is also convenient to enlarge the basis for the quark-gluon vertex to
include the total momentum of the meson.

\subsection{Transverse}\label{sec:bsetrans}
To avoid confusion, we will use different momentum labels here since our basic set of transverse orthogonal momenta are different to those in the previous section.  
To exploit the transversality of Landau gauge, we introduce the following 
transverse projections of the momenta
\begin{align}
\Omega^\mu_{T}        &= T_{\mu\nu}^{(\Delta)}\Omega^\nu\;,\\
\Sigma^\mu_{TT}       &= T_{\mu\alpha}^{(\Delta)} T_{\alpha\nu}^{(\Omega_T)} \Sigma^\nu\;.
\end{align}
Given explicitly, their normalised form is
\begin{align}
\hat{\Delta}^\mu      &= \left(\begin{array}{cccc} 0, & w^\prime y^\prime, & w y^\prime, & y\end{array}\right)\;, \\
\hat{\Omega}^\mu_{T}  &= \left(\begin{array}{cccc} 0, & -w^\prime y, & -w y, & y^\prime\end{array}\right)\;,\\
\hat{\Sigma}^\mu_{TT} &= \left(\begin{array}{cccc} 0, & -w, & w^\prime, & 0 \end{array}\right)\;,
\end{align}
with the angles $z$, $y$, and $w$ defined
\begin{align}
	\hat{\Delta}_{\perp\Omega}\cdot \hat{\Sigma}_{\perp\Omega} =w\;,\\
	\hat{\Delta}\cdot\hat{\Omega} = y\;,\\
	\hat{\Sigma}\cdot\hat{\Omega} = z\;,
\end{align}
where $\Delta^\mu_{\perp\Omega} = T_{\mu\nu}^{(\Omega)} \Delta^\nu$, $\Sigma^\mu_{\perp\Omega} = T_{\mu\nu}^{(\Omega)} \Sigma^\nu$ and
the hat indicates normalisation.  As before, $x^\prime=\sqrt{1-x^2}$. A suitable transverse and orthogonal basis is thus provided by
\begin{align}
&\left(
\begin{array}{cccc}
\cellcolor{blue!20}\gamma^\mu_{TTT}, &\cellcolor{blue!20} \hat{\Sigma}^\mu_{TT}, & \cellcolor{blue!20}\hat{\Omega}^\mu_T, & \hat{\Delta}^\mu\\
\end{array}
\right) \times R^{i}_{(\ONE)}\;,
\end{align}
where we define the Dirac part, to be used later, as
\begin{align}
\begin{array}{c}R^{i}_{(\ONE)}=\phantom{\Big\{}\\\phantom{\Big\{}
\end{array}
\begin{array}{cccccc}\Big\{ &
1,                                      & \hat{\slashed{\Delta}},                         & \hat{\slashed{\Sigma}}_{TT},                    & \hat{\slashed{\Omega}}_T,& \\ 
&\hat{\slashed{\Delta}}\hat{\slashed{\Sigma}}_{TT},&  \hat{\slashed{\Delta}}\hat{\slashed{\Omega}}_T,     & \hat{\slashed{\Sigma}}_{TT}\hat{\slashed{\Omega}}_T, & \hat{\slashed{\Delta}}\hat{\slashed{\Sigma}}_{TT}\hat{\slashed{\Omega}}_T &\Big\} \;. \\ 
\end{array}
\end{align}

Only those that remain under a transverse projection with respect to an external gluon
of momentum $\Delta^\mu$ are highlighted. This constitutes $24$ (reduced from $32$) components that allow for a convenient
representation of the quark-gluon vertex as needed in BSE calculations. It is here a function of three squared momenta
and three angles, with one squared momentum ($\Omega^2$) that corresponds to the mass of the bound-state considered as an external parameter in calculations.

Here, the triply transverse Dirac gamma matrix is defined
\begin{align}
\gamma^\mu_{TTT} &= T_{\mu\alpha}^{(\Delta)} T_{\alpha\beta}^{(\Omega_T)}T_{\beta\nu}^{(\Sigma_{TT})}\gamma^\nu\;, \nonumber \\
                 & = \left( \begin{array}{cccc} \gamma^1, & 0, & 0, & 0 \end{array} \right)\;.
\end{align}
As before, we may introduce the transverse momentum $v^\mu = \left(\begin{array}{cccc}1,&0,&0,&0\end{array}\right)$
as the orthogonal complement to $\Delta$, $\Omega$ and $\Sigma$ and thus define $\gamma^\mu_{TTT} = \slashed{v}v^\mu$.

We separate the Lorentz from the Dirac parts through
\begin{align}
L_{(H)}^\mu &= H^\mu\,,\;\;\ \quad \qquad\textrm{for~}H=\left\{ v,\hat{\Sigma}_{TT},\hat{\Omega}_T,\hat{\Delta} \right\} \,,\\
R^{i}_{(H)} &= \left\{ 
	\begin{array}{ ll}
		\phantom{\slashed{H}}R^{i}_{(\ONE)}
			& \quad \textrm{for~}H =\left\{\hat{\Sigma}_{TT},\hat{\Omega}_T,\hat{\Delta}\right\}\;, \\ 
		\slashed{H}R^{i}_{(\ONE)}
			& \quad \textrm{for~}H =\left\{v \right\}\;.
	\end{array}
	\right.
\end{align}

As before, the reduced quark-gluon vertex, as a function of the relative momenta $\Omega$ and $\Sigma$ and the total momentum  $\Delta$, can be written
\begin{align}
\Gamma^\mu(\Sigma,\Delta;\Omega) = \sum_{i=1}^8\sum_{a=\{v,\hat{\Sigma}_{TT},\hat{\Omega}_T\}}c_{a}^{i}(\Sigma,\Delta;\Omega) L^\mu_{(a)}R_{(a)}^i\;,
\end{align}
where the number of basis elements over which summation occurs is enlarged and we drop $a=\hat{\Delta}$ due to transversality.
The projectors, defined $\bar{R}^i$ and $\bar{L}_a^\mu$, satisfy
\begin{align}
\Tr\left[ \bar{R}^i_{(a)} R^{i^\prime}_{(a^\prime)}\right] &= \delta_{aa^\prime}\delta_{ii^\prime}\,,\qquad
 \bar{L}^\mu_{(a)} L^{\mu}_{(a^\prime)} &= \delta_{aa^\prime}\,,
\end{align}
such that
\begin{align}\label{eqn:basisprojection2}
c_{(a)}^{i}(\Sigma,\Delta;\Omega) = \bar{L}^\mu_{(a)} \Tr\left[ \bar{R}^i_{(a)} \Gamma^\mu(\Sigma,\Delta;\Omega)\right]\;.
\end{align}
where no summation on $a$ is implied on the right-hand side.

\subsection{Non-transverse}\label{sec:bsenontrans}
As an alternative to the basis above, we detail the obvious non-transverse basis that,
despite requiring the full $32$ elements of the extended basis to describe the
vertex, provides for a simple means to tackle the calculation.

Consider the dressed quark-gluon vertex in the BSE. A typical
choice for the momenta would be
\begin{align}
P^\mu &=  \ii\,M\,          \left(\begin{array}{cccc} 0, & 0,                 & 0,          & 1\end{array}\right)\,, \\
p^\mu &=  \left|p\right|\,  \left(\begin{array}{cccc} 0, & 0,                 & z^\prime,   & z\end{array}\right)\,, \\
k^\mu &=  \left|k\right|    \left(\begin{array}{cccc} 0, & w^\prime y^\prime, & w y^\prime, & y\end{array}\right)\,,
\end{align}
for the total momentum $P$, relative quark momentum $p$ and the internal loop momentum $k$.
From these, it is convenient to construct the following orthonormal momenta
\begin{align}
t^\mu &= \hat{P}^\mu\,, \nonumber\\
s^\mu &= \widehat{T^{(t)}_{\mu\nu}p^\nu}\,, \nonumber\\
r^\mu &= \widehat{T^{(t)}_{\mu\alpha}T^{(s)}_{\alpha\nu}k^\nu}\,, \nonumber\\
\gamma_{TTT}^\mu &:=T^{(t)}_{\mu\alpha}T^{(s)}_{\alpha\beta}T^{(r)}_{\beta\nu}\gamma^\nu = \gamma^\mu -\slashed{t}\,t^\mu- \slashed{s}\,s^\mu- \slashed{r}\,r^\mu\;.
\end{align}
We introduce the following orthonormal basis
\begin{align}
&\left(
\begin{array}{cccc}
\cellcolor{blue!20}\gamma^\mu_{TTT}, & \cellcolor{blue!20}r^\mu, & \cellcolor{blue!20}s^\mu, & \cellcolor{blue!20}t^\mu\\
\end{array}
\right) 
\times
\left(
\begin{array}{cccccccc}
1, & \slashed{r}, &\slashed{s}, & \slashed{t}, &\slashed{r}\slashed{s}, &\slashed{r}\slashed{t}, &\slashed{s}\slashed{t},&\slashed{r}\slashed{s}\slashed{t}   \\
\end{array}
\right)\;.\label{eqn:bsediracnontrans}
\end{align}
Here, we shaded the relevant basis elements in Landau gauge; this is all of them, since we do not exploit
transversality due to contractions with the ubiquitous gluon propagator.

Introducing $v$, the orthogonal complement to $r$, $s$, $t$, whose values for this specific frame given in Eq.~(\ref{eqn:ortho}) we can write
$\gamma_{TTT}^\mu = \slashed{v}v^\mu$ in order to separate the Lorentz from the Dirac part of the vertex.
We define the Dirac part of Eq.~(\ref{eqn:bsediracnontrans})
\begin{align}
R^{i}_{(\ONE)} =
\left\{\begin{array}{cccccccc}
1, & \slashed{r}, &\slashed{s}, & \slashed{t}, &\slashed{r}\slashed{s}, &\slashed{r}\slashed{t}, &\slashed{s}\slashed{t},&\slashed{r}\slashed{s}\slashed{t}\\
\end{array}\right\}\;,
\end{align}
and can then separate the Dirac from the Lorentz parts
\begin{align}
L_{(H)}^\mu &= H^\mu\,,\;\;\, \quad\;\quad\textrm{for~}H=\left\{ v,r,s,t \right\} \,,\\
R^{i}_{(H)} &= \left\{ 
	\begin{array}{ ll}
		\phantom{\slashed{H}}R^{i}_{(\ONE)}			& \quad \textrm{for~}H =\left\{r,s,t\right\}\;, \\
		\slashed{H}R^{i}_{(\ONE)}       			    & \quad \textrm{for~}H =\left\{v\right\}\;. 
	\end{array}
	\right.
\end{align}
As before, the reduced quark-gluon vertex is written
\begin{align}
\Gamma^\mu(p_1, p_2) = \sum_{i=1}^8\sum_{a=\{v,r,s,t\}}c_{a}^{i}(p_1,p_2) L^\mu_{(a)}R_{(a)}^i\;,
\end{align}
where the number of basis elements over which summation occurs is enlarged.
The projectors, defined $\bar{R}^i$ and $\bar{L}_a^\mu$, satisfy
\begin{align}
\Tr\left[ \bar{R}^i_{(a)} R^{i^\prime}_{(a^\prime)}\right] &= \delta_{aa^\prime}\delta_{ii^\prime}\,,\qquad
 \bar{L}^\mu_{(a)} L^{\mu}_{(a^\prime)} &= \delta_{aa^\prime}\,,
\end{align}
such that
\begin{align}\label{eqn:basisprojection3}
c_{(a)}^{i}(p_1,p_2) = \bar{L}^\mu_{(a)} \Tr\left[ \bar{R}^i_{(a)} \Gamma^\mu(p_1,p_2)\right]\;.
\end{align}
with no summation over $a$ implied.

\newpage

\end{document}